 \documentclass[prc]{revtex4}
\usepackage{graphicx} 
\usepackage{graphicx}
\usepackage{dcolumn}
\usepackage[utf8]{inputenc}
\usepackage{amsmath}
\usepackage{appendix}
\newcommand{\sect}[1]{\section{#1}\setcounter{equation}{0}}

\newcommand{\be}{\begin{equation}}
\newcommand{\ee}{\end{equation}}
\newcommand{\nn}{\nonumber}
\newcommand{\s}[1]{{\bf#1}}                
\def\to{\rightarrow}
\def\tofro{\leftrightarrow}
\def\half{\frac{1}{2}}

\def\tilde{\widetilde}

 \def\eqalign#1{%
\null\,\vcenter{\openup\jot\m@th
  \ialign{\strut\hfil$\displaystyle{##}$&$\displaystyle{{}##}$\hfil
      \crcr#1\crcr}}\,}
\def\no		{\nonumber}         
\def\hh		{\hspace{5 mm}}   
\def\lt		{\left(}                    
\def\rt		{\right)}                  
\begin{document}
\title{General Relativity {\it versus} Dark Matter for rotating galaxies}
\author{Yogendra Srivastava}\email{yogendra.srivastava@gmail.com}
\affiliation{Emeritus Professor of Physics, Northeastern University, Boston, MASS USA\\
\&\\
Dipartimento di Fisica e Geologia, Universit\'a di Perugia, Perugia, Italy}
\author{Giorgio Immirzi}\email{giorgio.immirzi@gmail.com}
\affiliation{Retired, Dipartimento di Fisica e Geologia, Universit\'a di Perugia, Perugia, Italy}
\author{John Swain}\email{jswain02115@yahoo.com}
\affiliation{Physics Department, Northeastern University, Boston MA USA}
\author{Orlando Panella}\email{orlando.panelle@pg.infn.it}
\affiliation{Istituto Nazionale di Fisica Nucleare, INFN Sezione di Perugia, Perugia, Italy}
\author{Simone Pacetti}\email{simone.pacetti@unipg.it}
\affiliation{Dipartimento di Fisica e Geologia, Universit\'a di Perugia, Perugia, Italy}
\begin{abstract}
A very general class of axially-symmetric metrics in general relativity (GR) that includes rotations 
is used to discuss the dynamics of rotationally-supported galaxies. The exact vacuum solutions of the 
Einstein equations for this extended Weyl class of metrics allow us to deduce rigorously the following:
(i) GR rotational velocity always exceeds the Newtonian velocity (thanks to Lenz's law in GR); 
(ii) A non-vanishing intrinsic angular momentum ($J$) for a galaxy demands the asymptotic constancy 
 of the Weyl (vectorial) length parameter ($a$) -a behavior identical to that found for the Kerr metric; 
(iii) Asymptotic constancy of the same parameter $a$ also demands a plateau in the rotational velocity. 
Unlike the Kerr metric, the extended Weyl metric can and has been continued within the galaxy and 
it has been shown under what conditions Gau\ss\ \&\ Amp\'ere laws emerge along with Ludwig's 
extended GEM theory with its attendant non-linear rate equations for the velocity field. Better estimates 
(than that from the Newtonian theory) for the escape velocity of the Sun 
and a reasonable rotation curve \&\ $J$ for our own galaxy has
been presented. 
\end{abstract}

\maketitle

\section{\bf Introduction \label{intro}} 
\noindent
The velocities of the ionized gases circling  many galaxies, as a function of the distance from
their centre, (the {\it rotation curves}),  do not appear to follow a Kepler law and drop as $1/\sqrt{r}$, 
but on the contrary tend to reach a constant plateau velocity ($v_\varphi$). This experimental fact, first
discovered by Vera Rubin\cite{VR} in the `80-s and confirmed by many later observations, poses one 
of the most interesting theoretical questions of today's physics. The most common explanation is 
to suppose that the observed mass and radius of a galaxy is only a small part of the total; the rest
being a vast (spherically symmetric) distribution of hypothetical dark matter (DM), which interacts 
only through gravitation; this is the basis of the widely accepted  $\Lambda$CDM model\cite{PB} 
(cosmological constant plus cold dark matter).\\
\\
The DM model when applied to (rotating) galaxies has its problems. First of all, that in spite of extensive searches 
no trace of this mysterious dark matter has been found. Secondly, there is an empirical but 
successful relation, the  Opik-Tully-Fischer law\cite{Opik:1922,Tully:1977,McG:2016}, between the plateau velocity of the gas ($v_\varphi$) and the visible -hence baryonic- mass of the galaxy: 
($M_{baryonic}\propto v_{\varphi}^4$).  But if the baryonic mass is supposed to be only a few percent of the total, how come this tiny fraction determines the rotational velocity of the galaxy? (Recall that in DM, an asymptotic $v_\varphi$ 
is generated by the {\it dark mass} not the baryonic mass). Thirdly, 
there has been no satisfactory explanation offered -in DM- for the magnitude of the observed angular momentum
($J_z$) of a galaxy. By contrast, in general relativity (GR), we can compute $J_z$ in terms of the rotation
velocity and the baryonic mass-current density that only extends over the visible size of any galaxy\cite{Weinberg:1972}. 
In fact, in a later section, we apply GR to the Milky Way and obtain  
a satisfactory 
estimate of $J_z$ for our own galaxy.   
\\
What we propose and show in this paper, building on previous work by other authors\cite{BG,Crosta,Cooperstock,Gross,Cornejo,Ludwig:1,Ludwig:2,Ludwig:3},
 that general relativity, when appropriately
applied, is perfectly capable of explaining the observed
phenomena above, provided one takes into account the finite size (and a non-spherical mass distribution) of most galaxies and the basic fact that they rotate.\\ 
To be concrete, let us consider our own galaxy \cite{Allen}. The Milky Way has a diameter of 25 Kilo parsec and a thickness of 2 Kilo parsec with a visible baryonic mass of about $(1\div2.5)\times 10^{11} M_\odot$. 
The considerably non-spherical geometry fixes the (stable) axis of rotation and our galaxy acquires a
 rotational velocity of about 200 km/sec at the edge (of the diameter). Rotations bring about a well-known but
  oft forgotten fundamental difference between the Newtonian theory \& GR.\\ 
\\
In the Newtonian theory, {\it there is no dependence of the gravitational field 
upon the rotation of a body}\cite{Stephani:1990}. In GR, on the other hand, 
the rotation of a system makes the metric non diagonal (i.e., the time-space 
component $g_{oi}\propto A_i$ becomes non-zero and a 3-vector-field $A_i$ is generated).
 A {\it preferred} direction (in space) is thus chosen and the {\it sense} of rotation (clock-wise or anti-clockwise) 
  established and fixed. This leads to the introduction of parity ($\mathcal{P}$) and 
  time-reversal($\mathcal{T}$) -violating but ($\mathcal{PT}$) conserving terms. 
  Thus, a geo-magnetic field ${\bf B}=\nabla\wedge {\bf A}$ emerges (already at 
  the linearized level in GR) that gives rise to the GEM (geo-electromagnetic) theory of 
  Thirring \& Lense\cite{Thirring:1918,Pfister:2012,Lense:1918}. [The ensuing Lense-Thirring 
  effect has been beautifully confirmed experimentally in\cite{Ciufolini:2004}]. An angular momentum
   ${\bf J}$ is generated (through the non diagonal term). These issues are discussed in detail in later sections.  
\\
The paper is organized as follows. In Sec.(\ref{Weyl}), we anchor our formalism upon  the most general class of 
stationary, axially-symmetric metrics found by Weyl\cite{Weyl:1917,Weyl:1918}. In this section, 
we discuss the Einstein equations valid in the vacuum (i.e., outside the galaxy). In Sec.(\ref{matter}), 
we consider the choice of the matter energy-momentum density appropriate for a galaxy that is supported 
entirely by rotations with zero pressure. The nature of the solutions of the Einstein equations for the matter
within the galaxy are explored. In Sec.(\ref{lenz}), we highlight a key role that Lenz's law plays in always 
boosting the rotation velocity up. In Sec.(\ref{TF}), we continue our discussion of Ludwig's extended GEM theory 
arising from the exact Weyl type constraints.
The affinity between the Weyl class of metrics and the specialized Kerr metric is commented 
upon in Sec.(\ref{kerr}) and in particular the appearance of an angular momentum whose value is 
computed for both. It is important to note that the Schwarzschild metric has zero angular momentum 
simply because it is spherical and thus lacks a vector field fixing a direction in space.
A simple phenomenological analysis using an analytic, factorized mass density is applied to obtain
 the rotation velocity for our own galaxy and compared with experimental data in Sec.(\ref{phen}). 
 We also estimate the angular momentum $J_z$ of our galaxy. The paper is concluded in Sec.(\ref{conc}) 
 with a summary of results obtained, work in progress and future prospects.      

\section{\bf The Weyl metric  \label{Weyl}}
\noindent
We shall write the axially-symmetric Weyl metric for a cylindrically symmetric space-time\cite{LL:1965}, with  coordinates
 $(ct,\varphi,r,z)$, including explicitly the rotation term (see, for example\cite{Stephani:1990}):
\be
ds^2=-e^{2U}(c dt-a d\varphi)^2+e^{-2U}\rho^2d\varphi^2+e^{2\nu-2U}(d\rho^2+dz^2),\quad
\label{w1}\ee
\be g_{\mu\nu}=
\left(\begin{matrix}
-e^{2U}& e^{2U}a&0&0\cr
e^{2U}a&-e^{2U}a^2+e^{-2U}\rho^2&0&0\cr
 0&0&e^{2\nu-2U}&0\cr 0&0&0&e^{2\nu-2U}
\end{matrix}\right);\ 
g=\det g_{\mu\nu}=-e^{4\nu-4U}\rho^2;\nonumber\\
\ee
the inverse metric has the form:
\be 
\label{w2}
g^{\mu\nu}=\left(\begin{matrix}
\frac{e^{2U}a^2}{\rho^2}-e^{-2U}&\frac{e^{2U}a}{\rho^2}&0&0\cr
\frac{e^{2U}a}{\rho^2}&\frac{e^{2U}}{\rho^2}&0&0\cr
0&0&e^{2U-2\nu}&0\cr 
0&0&0&e^{2U-2\nu}
\end{matrix}\right) ;\nonumber\\
 \ee
 and the invariant (spatial) volume element reads
 \begin{eqnarray}
 \label{w3}
 dV=(d\rho)(d\varphi)(dz)\sqrt{-g}= e^{-2(U-\nu)} (\rho d\rho dz d\varphi);\nonumber\\
 dV \geq dV_{flat}
 \end{eqnarray}
Below, we list some salient aspects of the above axially-symmetric metric:\\
\begin{itemize} 
\item 1: $U, a, \&\ \nu$ are functions only of $\rho=\sqrt{x^2+y^2}$ and $z$. independent of $\varphi$. 
Hence, there are two Killing vectors; one time-like and the other space-like (outside of the horizon) of the system. 
\item 2: The function $U$ is related to the Newtonian potential $\Phi$ through $e^{2U}=1+2(\frac{\Phi}{c^2})$.
\item 3: The function $a$ would be related to the angular momentum of the system. 
\item 4:  The gravito-magnetic potential-field $A_\phi=\frac{ca}{\rho}$, is a vector potential 
$\s A=(0, \frac{ca}{\rho},0)$.
\item 5: The three potential fields ($U, a, \&\ \nu$) characterizing the metric are not all independent. 
The Einstein equations in the vacuum, that is outside the boundaries of a confined system 
such as a galaxy, impose the following {\it exact} non-linear differential constraints on these functions\cite{Stephani:1990}:  
\begin{eqnarray}
\label{w4}
R_{\mu\nu}=0;\ {\rm in\ the\ vacuum\ of\ the\ system\ implies}:\nonumber\\
\Big{[}\frac{\partial^2 U}{\partial \rho^2} + \frac{\partial U}{\rho\partial \rho} +\frac{\partial^2 U}{\partial z^2}\Big{]}=-\Big{(}\frac{e^{4U}}{2\rho^2}\Big{)} \Big{[}(\frac{\partial a}{\partial \rho})^2 +(\frac{\partial a}{\partial z})^2 \Big{]};(i)\nonumber\\
\frac{\partial}{\partial z}\Big{(}\frac{e^{4U}}{\rho}\frac{\partial a}{\partial z} \Big{)} +\frac{\partial}{\partial \rho}\Big{(}\frac{e^{4U}}{\rho}\frac{\partial a}{\partial \rho} \Big{)}=0;(ii)\nonumber\\
{\rm and}\ (\frac{\partial \nu}{\rho \partial \rho})=\Big{[}(\frac{\partial U}{\partial \rho})^2-(\frac{\partial U}{\partial z})^2\Big{]}-\Big{(}\frac{e^{4U}}{4\rho^2}\Big{)} \Big{[}(\frac{\partial a}{\partial \rho})^2 -(\frac{\partial a}{\partial z})^2 \Big{]} 
;(iii)\nonumber\\
(\frac{\partial \nu}{\rho \partial z})= 2(\frac{\partial U}{\partial \rho})(\frac{\partial U}{\partial z})-(\frac{e^{4U}}{2\rho^2})(\frac{\partial a}{\partial \rho})(\frac{\partial a}{\partial z});(iv)\nonumber\\
\end{eqnarray}
N.B.:  Since $U$ and $a$ begin at order $G$, $\nu$ begins at second order (i.e., is of order $G^2$).
 Once $U$ \&\ $a$ satisfy the top two equations relating them Eq,(\ref{w4}(i),(ii)), a solution for $\nu$ exists
 since the last two equations Eqs.(\ref{w4}(iii),(iv)) become the integrability conditions for it; 
  $\nu \to 0$ as $\rho \to 0$ for any z. 
\item 6: The inequality in Eq.(\ref{w3}) that tells us that the invariant spatial volume element 
is larger than its value in the flat-limit is useful for proving bounds on integrals of (positive definite) 
integrands, in gravitational asymptotic perturbation theory such as that developed  by 
Landau-Lifshitz\cite{LL:1965} \&\ by Weinberg\cite{Weinberg:1972}.
\item 7:  A test particle in this axially symmetric metric would have two constants of motion, that we 
shall indicate as $p_0=E/c$ for time translations,  $ p_\phi=J/c$ for rotational motion in the x-y plane. 
We shall write $E=\gamma mc^2$, or $E=mc^2+{\mathcal E}_{NR}$ to study the non-relativistic limit.
\end{itemize}
We now write the geodesic equation for a test particle of mass $m$ for the above metric. The simplest
 formalism that extends to a Riemannian space blessed with a metric is through the action principle. 
 Calling the action $S$, $m$ the mass and $\tau$ the proper time $\tau$, we have 
\begin{eqnarray}
\label{t1}
dS=-(mc^2)d\tau;\ (dS)^2= (mc)^2(c d \tau)^2;\nonumber\\
{\rm Let}\ p_\mu =(\frac{\partial S}{\partial x^\mu});\ {\rm Hamilton-Jacobi\ Eqn.\ implies}:\ \ 
g^{\mu\nu}(\frac{\partial S}{\partial x^\mu})
(\frac{\partial S}{\partial x^\nu})=-(mc)^2;\nonumber\\
{\rm We\ have}\ \  p_\mu p_\nu g^{\mu \nu}=-(mc)^2;\nonumber\\
\end{eqnarray}
As stated earlier, an axially symmetric system has two conserved quantities: the energy $E$ and the component of angular momentum $J_z$ say, for rotational motion in the xy-plane. Hence, the dependence on time-interval ($t$) and that on $\varphi$ can be prescribed as
\begin{eqnarray}
\label{t2}
S(ct;\rho;\varphi;z)= -Et + J \varphi +{\hat S}(\rho;z);\nonumber\\
-\frac{\partial S}{\partial ct} = E/c;\ \frac{\partial S}{\partial \varphi}= J;\ \frac{\partial S}{\partial \rho}= p_\rho;\ \frac{\partial S}{\partial z}= p_z;\nonumber\\  
\end{eqnarray} 
Hence, for the Weyl metric, we have
\begin{eqnarray}
\label{t3}
(mc)^2= (\frac{E}{c})^2[e^{-2U}-(\frac{a}{\rho})^2e^{2U}]-2(\frac{a}{\rho})(\frac{J}{\rho})(\frac{E}{c})e^{2U}
-(\frac{J}{\rho})^2e^{2U} -e^{2(U-\nu)}[p_\rho^2+p_z^2];\ \ (i)\nonumber\\
(\frac{E}{c})^2e^{-2U} -[(\frac{J}{\rho})+(\frac{a}{\rho})(\frac{E}{c})]^2e^{2U}=(mc)^2+e^{2(U-\nu)}[p_\rho^2+p_z^2];\ \  (ii)\nonumber\\
{\rm Or}:\ [\frac{E}{c}\{e^{-U}+\frac{a}{\rho}e^U\}+\frac{J}{\rho}e^U][\frac{E}{c}\{e^{-U}-\frac{a}{\rho}e^U\}-\frac{J}{\rho}e^U]
= (mc)^2+e^{2(U-\nu)}(p_\rho^2+p_z^2);\ \ (iii)\nonumber\\
\end{eqnarray} 
Let $E= mc^2 \gamma$ and as both $E$ \&\ $J$ are constants of motion, we can define a reduced (a-dimensional) angular momentum, i.e., angular momentum per unit energy per unit $\rho$ (the perpendicular distance or, the impact parameter):$j\equiv (Jc/E\rho)$;\ and through it a {\it rotational} velocity $v_\varphi\equiv (j c)$. Similarly, the rotational parameter $a$ from the metric, can be employed to define a {\it vector potential}: $A_\varphi\equiv (ca/\rho)$ that has the dimensions of a velocity. With these definitions, Eq.(\ref{t3};(ii)) reads: 
\begin{eqnarray}
\label{t4}
J= \rho (\frac{E}{c}) j;\ v_\varphi= (c j) ;a= \rho (\frac{A_\varphi}{c}); \pi_\varphi \equiv (v_\varphi + A_\phi);\nonumber\\
\gamma^2\big{[}e^{-2U} -(\frac{\pi_\varphi}{c})^2e^{2U}\big{]}= 1+e^{2(U-\nu)}\big{[}\frac{(p_\rho^2+p_z^2)}{(mc)^2}\big{]};\nonumber\\
\end{eqnarray}
For a galaxy supported totally by rotations along $\varphi$, that is the focus of this paper, we set $p_z=0$ \& $p_\rho=0$.
Then the above equation is reduced to
\begin{eqnarray}
\label{t5}
\gamma = \frac{1}{\sqrt{[e^{-2U}- (\pi_\varphi/c)^2 e^{2U}]}};\nonumber\\
{\rm Keeping\ leading\ terms\ only}:\ \gamma\approx\  \frac{1}{\sqrt{[1-2U- (\pi_\varphi/c)^2]}};\nonumber\\
{\rm Test\ particle\ energy}: E=\gamma (mc^2)\approx\ mc^2 + \mathcal{E}_{NR};\nonumber\\
\mathcal{E}_{NR} = m\Phi +\frac{m}{2}\pi_\varphi^2;\ \pi_\varphi= (v_\varphi+A_\varphi);(ii) \nonumber\\
\end{eqnarray}
Eq(\ref{t5}(ii)) shows clearly what the Newtonian theory leaves out that GR supplies: viz., the vector potential $A_\varphi$. that in turn generates the GEM magnetic field. The lack of the dynamics generated by mass current density in the Newtonian theory is a serious lacuna that has important consequences. We discuss one such important improvement that GR provides.\\
As $U<0$, the particle will remain bound so long as $|v_\varphi+A_\varphi|<\sqrt{-2\Phi}$ and not $v_\varphi<\sqrt{-2\Phi}$
(their values at the coordinates $\rho,z$ in question) as the Newtonian theory asserts.\\
This leads to the well known quandry when one computes -using Newtonian gravity- the escape velocity of our Sun were it to escape from our Galaxy. The mean rotational velocity of our Sun is about 220\ Km/sec and it is approximately 8.2 Kilo-parsec away from the center of our Galaxy. There is apparently very little (baryonic) mass beyond this distance. Thus, Newtonian theory for the Sun's escape velocity predicts $\sqrt{2}\times (220)\approx\ 310$ Km/sec\cite{RAVE:2007} 
in the vicinity of our Sun, experimental astrophysicists estimate the Sun's escape velocity to be between ($500\div 550$) Km/sec.\\ 
In GEM, by contrast, the escape velocity reads: $v_{escape} \approx\ -A_\varphi+\sqrt{-2\Phi}$. As we shall discuss later in more detail, Lenz's law (reminding us that all masses attract so that the GEM magnetic field obeys the {\it left hand rule}) forces us to have $A_\varphi<0$, thus boosting the escape velocity up [{\it vedi} Sec.(\ref{lenz})]. From the phenomenology of the Milky Way in Sec(\ref{phen}), we estimate the magnetic term to add about 200 Km/sec, thereby bringing the escape velocity much closer to its estimated experimental value. A quantitative analysis of this matter shall be presented in a later work.\\  
\\
Having delineated a few important aspects that distinguish GR from the Newtonian theory regarding the dynamics of a rotation-supported galaxy, let us return to a discussion of the exact Weyl constraints.\\  
\\
At first glance, Eqs.(\ref{w4}(i-iv) appear quite opaque and daunting, but they acquire a physically more appealing aspect through the following {\it dictionary} in terms of the GEM electric ${\bf E}$ \&\ magnetic ${\bf B}$ fields of order $G$, along with a higher order field $\hat{\bf B}$ that is of order $G^2$. They are defined as follows: 
\begin{eqnarray}
\label{w5}
\s E&=(E_\rho,0,E_z)=(-\frac{\partial \Phi}{\partial\rho},0,-\frac{\partial \Phi}{\partial z})=-\nabla\,\Phi;(i)\cr
\s B&=(B_\rho,0,B_z)=(-\frac{\partial A_\varphi}{\partial z},0,\frac{\partial A_\varphi}{\partial\rho})=\nabla\wedge\s A;(ii)\cr
\s{\hat B}&=(\hat B_\rho,0,\hat B_z)=(-\frac{1}{\rho}\frac{\partial\nu}{\partial z},0,\frac{1}{\rho}\frac{\partial\nu}{\partial \rho});(iii);\nonumber\\
{\rm Thus,\ we\ have}:\ \s{\hat B}^2=\frac{1}{\rho^2}(\nu_\rho^2+\nu_z^2);(iv);\nonumber\\
\&\ -\rho(\nabla\wedge\s{\hat B})_\varphi=\nu_{\rho\rho}+\nu_{zz}-\frac{1}{\rho}\nu_\rho;(v);\nonumber\\
\end{eqnarray}
Before considering the equations they obey, let us pause to say a few words about the genesis of the 
nomenclature in Eq.(\ref{w5}). This EM analogy was first noticed and Eqs.(\ref{w5}(i-ii)) were used by Thirring. 
His initial purpose was to compute the gravitational field inside a hollow rotating sphere (in linearized GR). Later with Lense, he extended the analysis of the effect of proper rotation of a central body on the motion of other celestial
bodies, which led to the discovery of the Lense-Thirring effect\cite{Ciufolini:2004}. In a set of three beautiful papers,
Ludwig\cite{Ludwig:1,Ludwig:2,Ludwig:3} has extended GEM by including additional field energy (that are second 
order in $G$) and obtained a closed set of non-linear equations for the rotational velocity ($v_\varphi$) in terms of 
the Newtonian velocity (via its acceleration) and the matter distribution within the galaxy. We shall return to discuss them in a later 
section and show that indeed they are reproduced in the appropriate limit.\\
\\
In terms of the field variables defined in Eq.(\ref{w5}), the Weyl equations -in the vacuum- given in Eq.(\ref{w4}) read:  
\begin{eqnarray}
\label{w6}
\nabla\cdot{\bf E}=-(\frac{2}{c^2})e^{-2U}{\bf E}^2 +(\frac{c^2}{2}) e^{6U} {\bf B}^2;(i);\nonumber\\
\nabla\wedge{\bf B}= -(\frac{4}{c^2}) ({\bf E}\wedge{\bf B});(ii);\nonumber\\
{\hat{\bf B}}_\rho= (\frac{\rho}{c^2})[E_z^2-E_\rho^2] + \frac{e^{4U}}{4}[B_\rho^2-B_z^2];(iii);\nonumber\\
{\hat{\bf B}}_z = 2(\frac{\rho}{c^2})(E_\rho E_z) +(\frac{e^{4U}}{2})(B_\rho B_z);(iv);\nonumber\\  
\end{eqnarray} 
Within the galaxy, the {\it Gau\ss\ law} in Eq.(\ref{w6}(i)) shall get the mass density term on the right-hand side 
(-$4\pi \rho_m$). Similarly the {\it Ampere\ law} in Eq.(\ref{w6}(ii)) shall get the mass current density ($-4\pi \rho_m v_\varphi$) when we continue the solution within the galaxy. On the other hand, Eqs(\ref{w6}(iii-iv)) remain valid both inside
and outside of the galaxy, due to our choice of the matter energy-momentum density as discussed later in Sec.
(\ref{matter}) in detail.\\
 The various exponentials in these expressions add on higher order polynomials in the Newtonian potential due to the non-linearity of GR. In all the four equations above, the quadratic terms in ${\bf E}$ \& ${\bf B}$ appear; these are easily interpretable as different components of the field energy-momentum density.\\
 \\
An attentive reader might wonder how  (\&\ why) one can possibly succeed in describing the dynamics of a spin-2 gravitational field in terms of just the GEM-electric and magnetic (spin-1 vector) fields? The answer to this question 
lies in the non-linearity of GR. Already at the second order (in G), there are constraints between the ${\bf E}$-field (whose longitudinal part is defined through the gradient of the Newtonian potential $\Phi$ and whose transverse part arises through the time derivative of the transverse part of the vector potential, $\partial {\bf A}_T/\partial t$) and there are constraints between them, {\it vedi}
Eqs.(\ref{w4}(i-ii)). Further on, at order $G^2$, a subsidiary field $\nu$ appears in the metric as well as in the equations of motion, that is completely constrained by the behavior of the GEM fields and the boundary condition that 
$\nu(\rho=0;z)\equiv 0$. Thus, in the far field region, once the origin is appropriately chosen, the gravitational field is limited to its two degrees of freedom and its multipole expansion beginning with the quadrupole. Not so, in the near field within or in the vicinity of the galaxy where both longitudinal and transverse fields are present with constraints between them playing a crucial role in limiting the dynamics, as the following discussion illustrates.   
\\
The assumption that there is no motion along the (radial) $\rho$-direction or along the z-direction, brings in constraints for the dynamical system. Weinberg's Eq.(9.12)\cite{Weinberg:1972} gives the following expression for a particle's (spatial) acceleration $\mathcal{A}^i$ ( $i=2,3,4$ with coordinates labeled as $x^\mu:(x^1=ct,x^2=\varphi;x^3=\rho;x^4=z$)
\begin{eqnarray}
\label{a1}
\mathcal{A}^i = -\Gamma^i_{1,1} - 2\Gamma^i_{1,j}(\frac{dx^j}{dt}) - \Gamma^i_{j,k}(\frac{dx^j}{dt})(\frac{dx^k}{dt})\nonumber\\
\ \ \ \ \ \ \ \ \ +(\frac{dx^i}{dt})[\Gamma^1_{1,1}+ 2\Gamma^1_{1,j}(\frac{dx^j}{dt})+ \Gamma^1_{j,k}(\frac{dx^j}{dt})(\frac{dx^k}{dt})];\nonumber\\
\end{eqnarray}
Assuming only circular motion (about the z-axis), we have non-vanishing velocity only along the $\varphi$-axis:
$d\varphi/dt=v/\rho$ and $dx^i/dt=0$ for $i=3,4$. Under this premise, also the accelerations along the 3- \& 4-axes must vanish: 
\begin{align}
\label{a2}
(i)&\ \mathcal{A}^\rho =-c^2e^{4U-2\nu}U_{,\rho}+ce^{4U-2\nu}(\frac{v}{\rho})[a_{,\rho}+2aU_{,\rho}]
-e^{-2\nu}(\frac{v}{\rho})^2[-\rho+e^{4U}aa_{,\rho}+\rho^2U_{,\rho}+e^{4U}a^2U_{,\rho}]=0;\cr
(ii)&\ \mathcal{A}^z = -c^2e^{4U-2\nu} U_{,z}+ce^{4U-2\nu}(\frac{v}{\rho})[a_{,z}+2aU_{,z}]-e^{-2\nu}(\frac{v}{\rho})^2[\rho^2U_{,z}+e^{4U}a^2U_{,z}+e^{4U}aa_{,z}]=0;\nonumber\\
\end{align}
Eqs(\ref{a2}) along with Eqs.(\ref{w4}(i,ii)) allow us to obtain an exact non-linear, first order differential equation for the velocity field $\beta(\rho,z=0)=v(\rho,z=0)/c$ on the equatorial plane in terms of the (normalized dimensionless) Newtonian (velocity squared) defined as usual $g(\rho)=(\rho/c^2)(\partial \Phi(\rho,o)/\partial \rho)$, where $\Phi(\rho,0)$ is the Newtonian potential in the equatorial plane. We relegate this rather complicated expression to Appendix A. Here we shall illustrate the strategy employed to derive the result valid to the lowest non-vanishing order. To the desired order of accuracy, Eqs(\ref{a2}), yield the following expressions for $a,\rho$ \& $a_{,z}$:    
\be
\label{a3}
\frac{a,\rho}{\rho}= -(\frac{\beta}{\rho}) +(\frac{1}{\beta}+\beta)(\frac{\Phi_{,\rho}}{c^2});\qquad
 \frac{a,z}{\rho}=  +(\frac{1}{\beta}+\beta)(\frac{\Phi_{,z}}{c^2});
\ee
We can thus eliminate $a_{,\rho};a_{,z}$ in Eq.[\ref{w4}(ii)], to obtain an expression for the second derivatives of $U$. To the desired order of accuracy:
\be
\big[ e^{4U}(\frac{1}{\beta}+\beta)U_{,z} \big]_{,z} + 
[e^{4U}\{-\frac{\beta}{\rho}+(\frac{1}{\beta}+\beta)U_{,\rho}\}]_{,\rho}=0;
\label{a4}\ee
 Keeping only terms linear in the $U$-field:
 \be
(\frac{1}{\beta}+\beta)  [U_{,\rho,\rho}+U_{,z,z}] = (\frac{1-\beta^2}{\beta^2})(\beta_{,z}U_{,z})
+(\frac{1-\beta^2}{\beta^2})(\beta_{,\rho} U_{,\rho}) -\frac{\beta}{\rho^2}+\frac{\beta_{,\rho}}{\rho};
\ee 
Thus:
\be
[U_{,\rho,\rho}+U_{,z,z}+(\frac{U_{,\rho}}{\rho})]=
 [\frac{1-\beta^2}{\beta(1+\beta^2)}](\beta_{,z}U_{,z})
-(\frac{\beta^2}{\rho^2(1+\beta^2)})+(\frac{\rho \beta_{,\rho}}{\rho^2})
[\frac{\beta^2+(1-\beta^2)g(\rho,z)}{\beta(1+\beta^2)}]
+\frac{g(\rho,z)}{\rho^2};(i)
\ee
According to Eq.[\ref{w4})(i)], lhs is of\ order $G^2$, outside the galaxy. {\rm Thus, to linear order in G, we have at $z=0$ upon using the up-down symmetry, for the rate of increase of $\beta(\rho)$ (outside the galaxy)
\begin{eqnarray}
\label{a5}
(\rho \frac{\partial \beta}{\partial \rho})=\beta [\frac{\beta^2-g(\rho)(1-\beta^2)}{\beta^2+g(\rho(1+\beta^2))}];\nonumber\\
\end{eqnarray}
Eq.(\ref{a5}) is of course only valid outside the galaxy. It agrees exactly with Ludwig's Eq.(4.13)\cite{Ludwig:1}
when his solution is continued to outside the galaxy where the matter density term $f=0$.\\
\\
It is easy to obtain the rate equation inside the galaxy (to linear order) upon including the matter
density term on the rhs of Eq.(\ref{w4}(i)). To lowest order, the (2-dimensional) Laplacian of U receives 
the matter field contribution ($4\pi G \rho_m$).
Explicitly, inside the galaxy, we have
\begin{eqnarray}
\label{a6}
\nabla^2U(\rho,z)= (\frac{4\pi G\rho_m(\rho,z)}{c^2})+{\rm terms\ of\ order}\ G^2;\nonumber\\
{\rm Define\ for}\ z=0;\ f(\rho)= (\frac{4\pi G\rho_m(\rho,z=0) \rho^2}{c^2});\nonumber\\
Eq.(\ref{a4}(i)) \to\ 
(f-g) +\frac{\beta^2}{1+\beta^2}=\frac{1}{\beta(1+\beta^2)}(\rho\frac{\partial \beta}{\partial \rho})[\beta^2
+g(1-\beta^2)];\nonumber\\
(\rho\frac{\partial \beta}{\partial \rho})= \beta \big{[}\frac{\beta^2+(1-\beta^2)(f-g)}{\beta^2+g(1+\beta^2)}\big{]};\nonumber\\ 
\end{eqnarray}
This essentially reproduces Ludwig's result inside the galaxy and reduces to Eq.(\ref{a5}) outside the galaxy for which $f=0$.
\section{\bf Matter energy-momentum density \label{matter}}
\noindent
Within the boundaries of the galaxy, the dynamics of course changes:
\begin{eqnarray}
\label{w7}
E_{\mu \nu}(\rho,z) = R_{\mu\nu} -(\frac{1}{2}) R\  g_{\mu\nu} = (\frac{8\pi G}{c^4}) T_{\mu\nu};\nonumber\\
\end{eqnarray}
and thus we need a model for the energy-momentum density of the rotating galaxy and a choice for the metric inside. 
Hoping that no confusion ensues, we shall continue to use the same form of the metric as given in Eq.(\ref{w1}).
The simplest and most commonly used model for matter is that of {\it free dust} with in general an equation of state relating the mass density to the pressure. We shall assume further that our galaxy has zero-pressure, which implies that it is {\it totally}
supported by rotations around its stable axis, with no further extraneous motion. Choosing the
axis of rotation along the z-axis (with an angular velocity $\dot{\varphi}$), our extreme simplifying  assumptions, allow us to restrict the matter energy-momentum density
to the following form [with coordinates ($o,\varphi,\rho,z$)]:
\begin{eqnarray}
\label{w8}
T^{\mu\nu}= \rho_m u^\mu u^\nu; \nonumber\\
u^\mu(\rho,z)= (\gamma c)(1,\frac{\beta}{\rho},0,0);\nonumber\\
u_o=-(\gamma c)e^{2U}[1- \beta\frac{a}{\rho}];\ u_\varphi=(\gamma c)\Big{[}e^{2U}a(1-\beta \frac{a}{\rho})+(\beta \rho) 
e^{-2U}\Big{]}; u_\rho=0; u_z=0; \nonumber\\
{\rm The\ trace}:\ T^\mu_\mu=-(\rho_m c^2) \Rightarrow\ [\frac{1}{\gamma^2}]= [(1-\beta\frac{a}{\rho})^2e^{2U}- 
\beta^2 e^{-2U}];\nonumber\\ 
\end{eqnarray}
While lack of motion along the $\rho$ (radial) \& \  z (vertical) directions simplify the structure of the matter energy-momentum density tensor from a ($4\times4$) matrix to a ($2\times2$) matrix form, this simplification also brings some unexpected peculiarities such as:
\begin{itemize} 
\item 1: Even though the reduced matrix $T_{\mu\nu}$ is real-hermitean, it is non-diagonal and because it is factorizable  its determinant is zero. We recall that in the general case, this matrix has 4 eigenvalues: a positive definite (time-like) mass density with 3 (space-like) pressures ($p_1,p_2, p_3$ along its principal axes). By setting all pressures $p_i$ to zero, we have made the matrix {\it singular} with the lone non-vanishing eigenvalue the scalar (generally invariant) mass density $\rho_m c^2$.
\item 2: For any finite $\beta$, the Lorentz factor $\gamma$ in Eq.(\ref{w8}) does not reduce to its expected value 
$(1-\beta^2)^{-1/2}$, unless the rotation parameter $a\to 0$. But, if we let $a=0$, the metric becomes {\it diagonal}, since then $g_{o\varphi}=0$ thereby rendering the (matter+field) angular-momentum zero. Clearly, this is 
unphysical and thus unacceptable. We must have $a\neq 0$ (it can be positive or negative, of course).
\item 3. In the expression for $\gamma$, the linear term in $\beta$ induced by a non-vanishing length parameter
$a\neq 0$, would exceed the expected $\beta^2$ correction unless {\it for any value of $\rho\leq \rho_{edge}$ within the galaxy}, $2|a(\rho)/\rho|<\beta(\rho)$. In short, $\beta$ can not be {\rm too small} if the rotational velocity alone has to support a galaxy with zero internal pressure.
\item 4: The metric and its first derivatives must be matched at the boundary for their inside versus outside values. 
\end{itemize}
Thus, $\beta$ just outside cannot be too small either. 
A clear indication from GR that Newtonian values for $\beta$ that are becoming too 
small at the edge must get supplemented by (the mass current density) contributions to 
stabilize the system.\\          
To emphasize the affinity and the difference between Einstein gravity and electromagnetism, 
and partly to follow the works by Ludwig\cite{Ludwig:1,Ludwig:2,Ludwig:3},
 it is convenient to write the Einstein equations for this metric in terms of the
three vectors $\s E, \s B, \s{\hat B}$ defined  earlier.
Overall  we have a dictionary with which we can write the Einstein equations
\be
\label{E1}
 E_{\mu \nu}\equiv R_{\mu \nu}-\half g_{\mu\nu} R=\frac{8\pi G}{c^4}T_{\mu\nu}
\ee
We have:
\begin{align}
\label{E2}
 R&=g^{\mu\nu}R_{\mu\nu}=\frac{8\pi G}{c^4}g^{\mu\nu}T_{\mu\nu}=e^{2U-2\nu}\big(2\nabla^2U+
\frac{e^{4U}}{\rho^2}(a_{,\rho}^2+a_{,z}^2)-2(\nu_{,\rho,\rho}+a_{,zz}+U^2_{\,rho}+U^2_{,z})\big)=\cr
&=e^{2U-2\nu}\big(-2\frac{e^{-2U}}{c^2}\nabla\cdot\s E-4\frac{e^{-4U}}{c^4}\s E^2+16\frac{e^{4U}}{c^2}\s B^2
+2\rho(\nabla\wedge\s{\hat B})_\varphi+\hat B_\rho\big)
\nn\end{align}
and therefore a ``Gau\ss\ law''
\be 
\label{E3}
\nabla\cdot\s E=-4\pi G\rho_m e^{2\nu}(1+e^{-2U}(\beta\gamma)^2)-2\frac{e^{-2U}}{c^2}\s E^2+8e^{6U}\s B^2
+\rho c^2e^{2U}(\nabla\wedge\s{\hat B})_\varphi-\half c^2e^{2U}\hat B_z
\ee
To single out the non-diagonal part of $ E_{\mu \nu}$ in terms of the matter current density
$\s J_m=\rho_m\s v_\varphi$, we consider the combination
\begin{align}
\label{E4} 
&a E_{ct\,ct}+E_{ct\,\varphi}=\frac{8\pi G}{c^2}(aT_{ct\,ct}+T_{ct\,\varphi})=
\frac{8\pi G}{c^2}(-(J_m)_\varphi \gamma^2\rho(1-\frac{a}{\rho}\beta))=\cr
&=-\half e^{4U-2\nu}\big(a_{,\rho,\rho}+a_{,z,z}-\frac{1}{\rho}a_{,\rho}+4(a_{,\rho}U_{,\rho}
+a_{,z}U_{,z})\big)=\cr
&= \frac{2\rho}{c} e^{4U-2\nu}\big((\nabla\wedge \s B)_\varphi-4(\s E\wedge\s B)_\varphi\Big)
\end{align}
and therefore an ``Amp\`ere law'' emerges:
\be
\label{E5}
\nabla\wedge\s B=\frac{4\pi G}{ c}e^{-4U+2\nu}(-\s J_m \gamma^2(1-\frac{a}{\rho}\beta))
+\frac{c}{2\rho}e^{-4U+2\nu}\s E\wedge\s B
\ee
For the convenience of the reader, in Appendix B, we have reproduced some details of the traditional iterative scheme in GR (developed over a century ago). Anyone interested can readily compare the higher order contributions as they arise from the perturbative scheme with the exact Einstein-Weyl equations.\\  
\\
Neglecting higher order term in $G$ and (special) relativistic corrections, we can summarize Gau\ss\ and Amp\`ere law as:
\be
\label{E6}
\nabla\cdot\s E=-4\pi G\rho_m,\qquad \nabla\wedge\s B=-\frac{4\pi G}{ c}\s J_m
\ee
It is important to note (and very useful to remember to implement) the negative sign of the matter fields on the rhs of Eqs.(\ref{E6}), especially in the Amp\'ere
law that leads to a {\it left hand rule} for the GEM magnetic field. Precisely because gravitation has only attraction (unlike E\&M that has both), the Lenz's law for gravity implies that there is a net boost to the acceleration due to other masses. 
We illustrate in Sec(\ref{lenz})  
that the model obeying Lenz's law produces a rotation velocity curve consistent with 
mass-to-luminosity data whereas another model while successful in producing the rotation curve was inconsistent with the light intensity data.\\ 
\\
\section{\bf Lenz's law always boosts rotational velocities for stable galaxies \label{lenz}}
An attentive reader might rightly wonder why there is always a counter rotating  GEM magnetic field produced 
by the velocity-field of material masses. Such is not always the case in Maxwellian electrodynamics due to the fact
that both attractive and repulsive forces are generated as both positive and negative charges exist in the electro-magnetic
theory of Maxwell. In GEM however, the force is always attractive\cite{Ruggiero:2002,Mashoon:2008}. For the problem
at hand, it is most easily seen in the equation for the GEM magnetic field   
\begin{eqnarray}
\label{L1}
\nabla \times {\bf B} = - (\frac{4\pi G}{c^2}) \rho {\bf v} + {\frac{\partial {\bf E}}{c^2\partial t}};\nonumber\\
\end{eqnarray}
The minus sign in the first term on the right hand side of Eq.(\ref{L1}) tells us that the magnetic field induced
on the left side (due to the velocity field) follows the {\it left-hand rule} always. In standard electrodynamics with different signs of charge, 
Lenz's law implies that a negatively charged electron in a  beam of co-moving electrons loses momentum due to other negatively charged electrons in the beam. On the other hand, the same Lenz's law implies that an electron gains momentum if there are say positively charged parallel moving protons. In GEM, there is only attraction between masses and thus the situation is similar to that between an electron and a proton. Ergo, Lenz's law implies that there is always an increase in the rotational velocity of galaxies due to GEM. In the following sections, we shall confirm these results explicitly that the resultant rotational velocity is indeed boosted through a GEM magnetic term $B_z<0$.\\
\begin{figure}
\scalebox {0.7}{\includegraphics{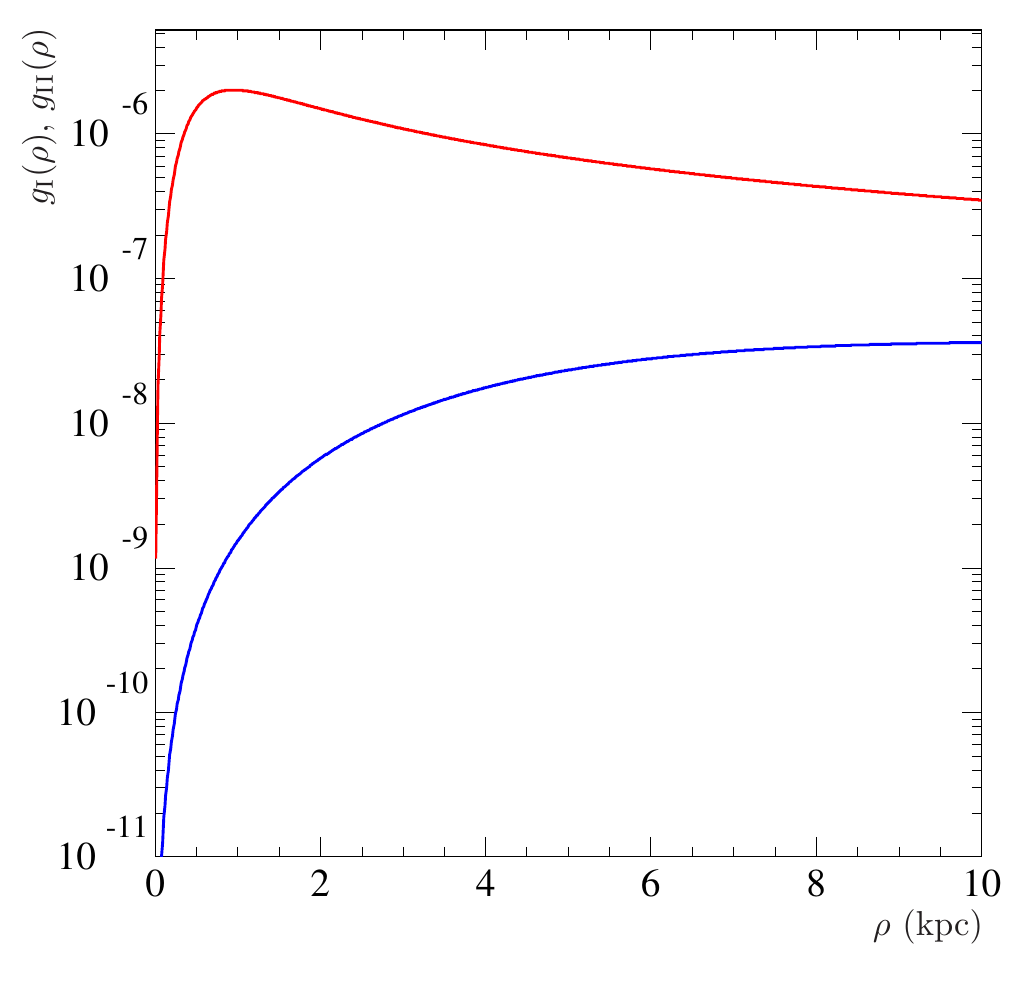}}
\caption{Newtonian g-functions for the two models as defined in Eq.(\ref{i3}) in the text are shown in this figure
with $g_I$ in red and $g_{II}$ in blue}
\label{Fig(1)}
\end{figure}
\begin{figure}
\scalebox {0.7}{\includegraphics{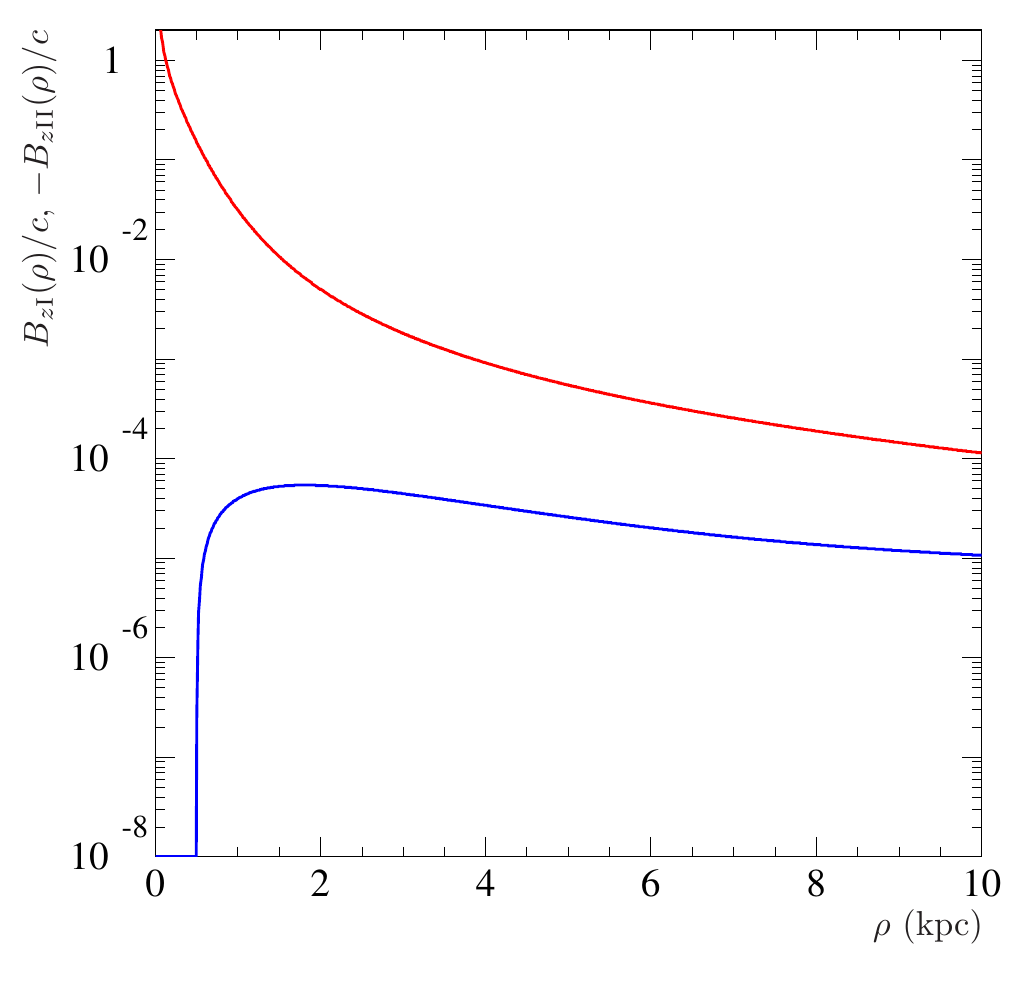}}
\caption{$B_z$ for model I \& $-B_z$ for model II are shown in this figure. Model I has the wrong sign while Model
II has the correct sign according to Lenz's law.}
\label{Fig(2)}
\end{figure}
\begin{figure}
\scalebox {0.7}{\includegraphics{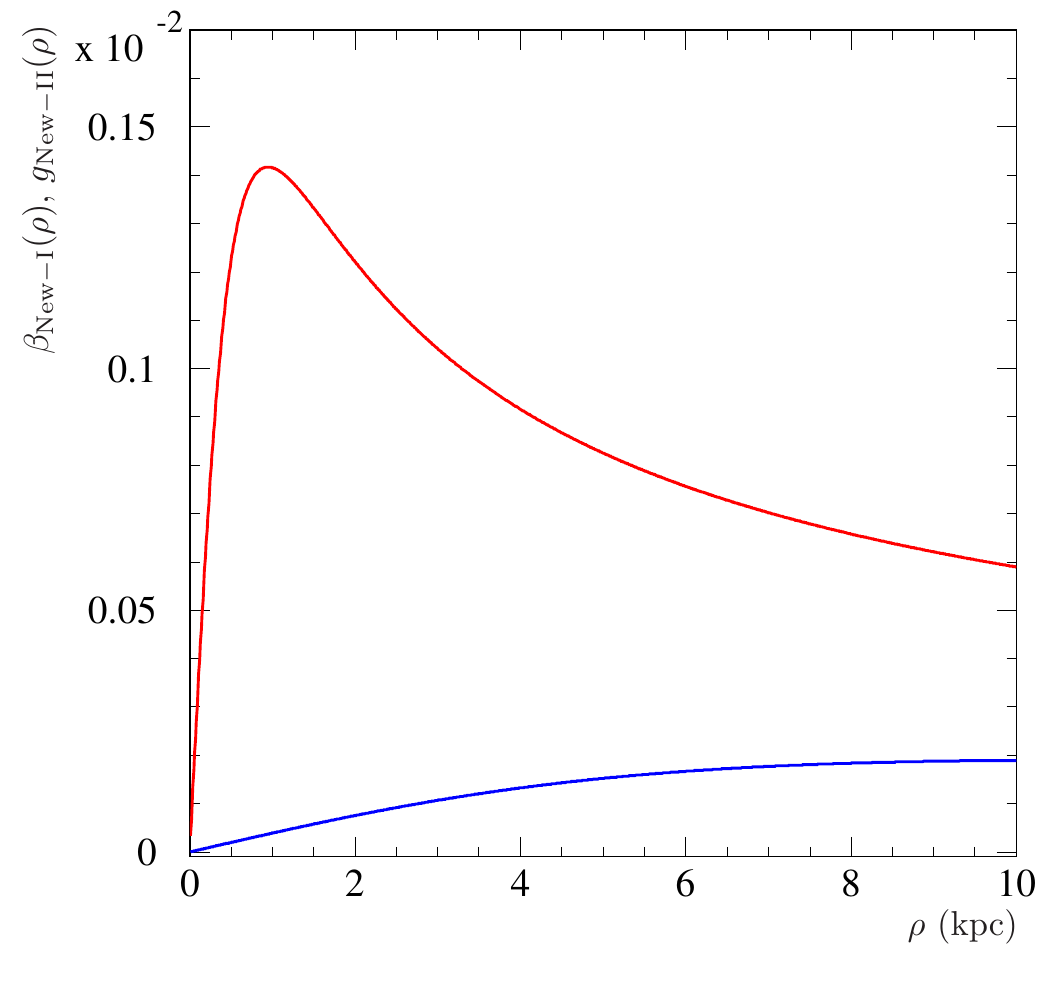}}
\caption{We show the normalized Newtonian velocities $\beta$-functions, $g_I$ in red and $g_{II}$ in blue.}
\label{Fig(3)}
\end{figure}
We discuss it below  
and show  
that the model obeying Lenz's law produces a rotation velocity curve consistent with 
mass-to-luminosity data whereas another model while successful in producing the rotation curve was inconsistent with the light intensity data.\\ 
The example of galaxy NGC 1560 has been discussed at length in \cite{Ludwig:1} using two different 
parametrisations, we shall call them model 1 \& model II:
\begin{eqnarray}
\label{i1}
{\rm model\ I:}\  R_s = 7\times 10^{-6}\ Kpc; a=0.373\ Kpc;\ b=0.300\ Kpc;\nonumber\\ 
{\rm normalization\ point}: \beta(8.29 Kpc) = 2.67\times 10^{-4};\nonumber\\
{\rm model\  II:}\  R_s = 1.46\times 10^{-6}\ Kpc; a= 7.19\ Kpc;\ b= 0.567\ Kpc;\nonumber\\
 {\rm normalization\ point}: \beta(8.29 Kpc) = 2.67\times 10^{-4};\nonumber\\
\end{eqnarray}
They both produce roughly the same $\beta(\rho)$.To illustrate our point as simply as possible, we made a simple interpolation of the numerical result that Ludwig found from his rate eqn. The interpolation reads 
\begin{equation}
\label{i2}
\beta(\rho) \approx\ (2.64\times 10^{-4})[\frac{\rho^2}{(\rho^2+2.92)}];\ ({\rm all\ distances\ in\ Kpc}). 
\end{equation}
The Newtonian g-functions for the two models are as follows:
\begin{eqnarray}
\label{i3}
g_I(\rho)= (3.5\times 10^{-6}) \big{[} \frac{\rho^2}{[\rho^2+ 0.45]^{3/2}} \big{]};\nonumber\\
g_{II}(\rho)= (7.3\times 10^{-7}) \big{[} \frac{\rho^2}{[\rho^2+ 60.17]^{3/2}} \big{]};\nonumber\\
\end{eqnarray}
These are shown in Fig.(1).
The GEM magnetic field is defined as
\begin{equation}
\label{i4}
\frac{B_z}{c} = \frac{g(\rho)-\beta^2}{\beta\rho}.
\end{equation}
For model I, $B_z>0$ and for model II, $B_z<0$. In Fig(2), we show the magnetic fields, $Bz_I$ for model I and
$-B_{z,II}$ for model II. Lenz's law is not obeyed in model I but it is in model II. In Fig.(3), we show the corresponding
Newtonian velocities\\
Ludwig's model II obeys Lenz's law and at the same time is also consistent with the mass-to-luminosity data, whereas model I does not agree with the mass-to-luminosity data. This shows the efficacy of Lenz's law in limiting the class of solutions.

\sect{\bf More on rotation velocity and the Tully-Fisher law \label{TF}}
\noindent
As discussed in Sec.(\ref{lenz}), the induced GEM magnetic field ${\bf B}$ is always counter-rotating
(follows the left hand rule) with respect to velocity-field of material masses that produce it. 
Also, as shown earlier, the Einstein-Weyl equations acquire the form of
 Gau\ss\ -like and  Amp\`ere-like  laws, even at the linearized level.\\ 
 Upon assuming that $\s A_g=A_\varphi  \hat\varphi;\ \ \s v=v\,\hat\varphi$ and that we are in stationary conditions, 
 the equations (in cylindrical coordinates) read\cite{Ludwig:1}:
\begin{eqnarray}
\label{T1}
\phi_g = (\frac{\Phi}{c^2});\nonumber\\
      \frac{1}{\rho}\frac{\partial}{\partial \rho}\big(r\frac{\partial\phi_g}{\partial \rho}\big)+
\frac{\partial^2\phi_g}{\partial z^2}=\nabla^2\phi_g=4\pi G\rho_m;\quad;\nonumber\\
 \frac{\partial}{\partial \rho}\big(\frac{1}{\rho}\frac{\partial (\rho A_\varphi)}{\partial \rho}\big)+
      \frac{\partial^2A_\varphi}{\partial z^2} =\frac{4\pi G}{c^2}\rho_m\,v ;\nonumber\\
 \end{eqnarray}
The assumption is  that $v(\rho,z)$ describes continuously the motion of the 
rotating matter inside the galaxy and the motion of the ionized gas that circles round it.
While the geodesic equations for the (spatial) acceleration  of a particle
$\mathcal{A}^i$ have been shown to be non-linear and complicated, however,
we want to limit our discussion here and consider only equatorial circular motion 
around the z-axis with $\frac{d\varphi}{dt}=\frac{v}{\rho}$
and $\mathcal{A}^\rho= \mathcal{A}^z=0$. Under these provisions, to lowest order we have
the Lorentz force equations:
\begin{eqnarray}
\label{T2}
\frac{\partial \Phi}{\partial\rho}-\frac{v^2}{\rho}=\frac{v}{\rho}\frac{\partial(ca)}{\partial\rho},\qquad
\frac{\partial \Phi}{\partial z}=\frac{v}{\rho}\frac{\partial(ca)}{\partial z}\quad\tofro\ 
 E_z-vB_\rho=0;\quad E_\rho+vB_z=-\frac{v^2}{\rho};\nonumber\\
 {\rm Define,\ a\ magnetic\ velocity\ term}:\ \beta_{mag}\equiv\ \frac{\rho (-B_z)}{c}\geq 0;\nonumber\\
 {\rm Thus, with\ g\ the\ Newtonian\ velocity\ squared}: \beta^2 = g + (\beta \beta_{mag}) \geq g;(i)\nonumber\\
 \beta = (\frac{1}{2}) \big{[} \beta_{mag}+\sqrt{(4g+\beta_{mag}^2)}\ \big{]};(ii)\nonumber\\
\end{eqnarray}
Thus, as we proposed to show in Sec (\ref{intro}),
GR with its inherent Lenz's law does indeed produce the remarkable result that 
the rotational velocity always exceeds its Newtonian value: [$\beta^2\geq g$ Eq.(\ref{T2}(i))].\\
To put it in perspective, this relationship is amply confirmed through 2700 data points from
153 SPARC galaxies. For details, we refer the reader to \cite{McG:2016}, especially Fig.(3)
in it.\\ 
We have also shown that up to the order of required accuracy, Ludwig's rate equations for the rotation
velocity emerge from the Weyl metric, thereby giving strong support to Ludwig's computational 
program. We shall return to it in Sec.(\ref{Ludwig}).\\   
\\
A simple qualitative argument for constant
asymptotic velocity can be deduced from these equations, with a Newtonian term augmented by the
magnetic term. At small distances from the center, the Newtonian term dominates but as one proceeds
further towards the edge of the galaxy, the picture changes dramatically due to the on set of the
magnetic term.\\ 
\\
If we consider our own galaxy, the Newtonian velocity has roughly speaking
two bumps and then it goes down in the Keplerian fashion as $1/\sqrt{\rho}$.
If we simply add a magnetic term that begins from zero and grows up near the edge to produce a constant
(negative) vector potential $A_\varphi$ in obedience to the Lenz's law,
we have the desired result of a constant rotational velocity, the by now well established
result, first found experimentally by Vera Rubin.\\
We also notice that the same asymptotically constant vector potential allows us to obtain
a reasonable estimate both for the rotation velocity \&\ the angular momentum of our galaxy.\\    
\\
For our galaxy, 
the maximum of the Newtonian term coincides approximately with the on set of asymptotic velocity,
$\beta^2(\infty)=(\frac{R_s}{2R_{edge}})$, where the Schwarzschild radius $R_s= (2GM/c^2)$ with $M$
denoting the baryonic mass (plus that of the gravitational field). For a pillbox like galaxy, 
$V= (\pi R_{edge}^2) h,\ M=(\rho_mV)$, so that $\beta^2(\infty)\sim (\frac{M}{M^{1/2}})\sim M^{1/2}$, 
reproducing the Tully-Fisher law: $M\propto\ \beta^4$.

\section{\bf Weyl class of metrics \&\ the particular Kerr metric \label{kerr}}

We wish to investigate the similarities and differences between 
the large distance behaviour of the Weyl class of metrics to the particular
one of the Kerr solution  of the Einstein equations\cite{Stephani:1990}. This solution apparently 
describes a rotating black hole in terms of a mass $M$ and a (constant) length parameter $a$ that 
is known to be linearly related to its angular momentum.\\ 
Taking  $\hat z$ as axis of rotation, $g_{\mu\nu}\equiv\eta_{\mu\nu}+h_{\mu\nu}$, 
 at large distance, the Kerr metric asymptotic  behaviour  is given by (\cite{Weinberg:1972} pg. 240):
\be
h_{ij}\to -\frac{R_s}{r^3}x_ix_j, \quad h_{0i}\to \frac{R_s}{r^2}(x_i+\frac{1}{r}(\s a\wedge \s x)_i),\qquad
R_s\equiv\frac{2GM}{c^2}.\ \s a=(0,0,a),\ i,j=1,2,3
\label{asymK}\ee 
As amply discussed in Appendix(\ref{iter}), this  is quite generally all that one needs
to calculate the total mass and angular momentum.   
For the Kerr metric\eqref{asymK}, Eq.(\ref{II-2}) yields $ E_{tot}= Mc^2,\ \s J= (Mc)\s a$, as expected. If $a=0$ the 
Kerr metric coincides with the Schwarzschild metric and $\s J$ is zero. 
We can see that  for the system to have a finite angular momentum, and a rotating 
galaxy certainly has that, it is crucial that the space-time part of $h_{\mu\nu}$ does not vanish asymptotically beyond $1/r^2$.\\
Let us now consider the general class of Weyl's axially-symmetric metrics as in 
Sec(\ref{Weyl}) focusing on their space-time part in the equatorial plane (i.e., at $z=0$ so that $\rho=r$) and we have:   
\begin{eqnarray}
\label{E3}
g_{o\varphi}(r)= \frac{a(r)}{c}e^{2U(r)},
 {\rm can\ be\ written\ in\ pseudo-Euclidean\ coordinates as\ the\ special\ case\ of}\ g_{oi}=
\epsilon_{ijk} a_j x_k (\frac{e^{2U}}{r^2});\nonumber\\
{\rm with\ Weyl's\ being\ the\ special\ case}\ {\bf a}= (0,0,a);\nonumber\\
{\rm Expanding\ in\ perturbation\ theory:}\nonumber\\  
g_{oi}=g_{oi}^{(1)}+g_{oi}^{(2)}=\epsilon_{ijk}\frac{a_jx_k}{r^2} [1+2U(r)+....],\nonumber\\
{\rm with}\  g_{oi}^{(1)}= \epsilon_{ijk}\frac{a_jx_k}{r^2}\ ;\&\  g_{oi}^{(2)}= \epsilon_{ijk}\frac{a_jx_k}{r^2}(2U(r);\nonumber\\
\end{eqnarray}
We are interested in the second part ($g_{oi}^{(2)}$) that relates 
to the angular momentum (${\bf J}$) of the system. 
Asymptotically, we have (vedi, \cite{Weinberg:1972}) for the second term,
\begin{eqnarray}
\label{E5}
g_{oi}^{(2)}= (\frac{2G}{r^3})(\bf{r}\times\bf{J})_i;\nonumber\\
{\rm Using\ Eq(\ref{II-2})\ we\ find}\  J_z= (Mc) a; 
\end{eqnarray}  
exactly the same as that for the Kerr metric provided we associate the (constant) Kerr length 
parameter $a$ with the (asymptotic) Weyl length parameter $a$.\\
The implication is that a finite value of the total (material+that of the gravitational field) angular momentum 
of the galaxy requires that the rotational velocity asymptote to a constant value and {\it vice versa}.\\
A mental picture of what is happening may be formed through the following rough guide about the 
Weyl parameter $a$. For small $r$, $a$ increases from zero linearly until the edge, beyond which -while
continuous at the edge- it eventually becomes a constant. At very large $r$, as expected the GEM 
magnetic field  ($-B_z\to 1/r$), as all radiation fields do. 

\section{\bf Ludwig's non-linear differential equation for the velocity field \label{Ludwig}}

While in Sec.(\ref{TF}) Eq.(\ref{T2}) we have tried to keep our equations {\it linear} by keeping both
the Newtonian and the magnetic contributions at the same level, the strategy followed by
Ludwig\cite{Ludwig:1}(see also\cite{Ludwig:4,Ludwig:5}) has been to eliminate the magnetic term 
entirely, at the expense of course of ending up with a non-linear equation for the velocity field. Below 
we follow his formalism to pinpoint a few aspects.\\ 
As stated in the last paragraph, we can use Eq.(\ref{T1}) to eliminate $A_\varphi$ from 
the expression of the Amp\`ere law, that becomes
   \be 
    \frac{\partial}{\partial \rho}(\frac{1}{v}\frac{\partial\phi}{\partial \rho}-\frac{v}{\rho})+
    \frac{\partial}{\partial z}(\frac{1}{v}\frac{\partial\phi}{\partial z})=\frac{4\pi G}{c^2}\rho_m v.
    \label{L1}\ee
 This equation  multiplied by $v$ and subtracted from the expression of Gau\ss' law given earlier,
  eliminates the double derivatives and yields:
\be   
4\pi G\rho_m(1-\frac{v^2}{c^2})=(\frac{1}{\rho}+\frac{1}{v}
\frac{\partial v}{\partial \rho})\frac{\partial\phi_g}{\partial \rho} +
\frac{1}{v}\frac{\partial v}{\partial z}\frac{\partial\phi_g}{\partial z} +
v\frac{\partial}{\partial \rho}\frac{v}{\rho}
\label{L2}\ee
 a non linear first order differential equation for $v(\rho,z) $ for given $\rho(\rho,z)_m,\ 
   \Phi_g(\rho,z)$. In the equatorial plane $z=0$ by the up-down symmetry
   we can drop the $\frac{\partial\phi_g}{\partial z}$; then:
   \be
   (\beta^2+\rho\frac{\partial\varphi}{\partial \rho})r\frac{\partial\beta}{\partial \rho}=
   \frac{\beta}{\rho}\big((\beta^2-r\frac{\partial\varphi}{\partial \rho})+\frac{4\pi G\rho_m}{c^2}
   \rho^2(1-\beta^2)\big);\quad \beta=\frac{v(\rho,0)}{c},  \varphi=\frac{\phi_g}{c^2}
   \nn\ee
Outside the galaxy, where $\rho(\rho,0)_m=0$, the equation becomes
\be
  \frac{\rho}{\beta}\,       \frac{\partial\beta}{\partial \rho}=\frac{\beta^2-\rho\frac{\partial\varphi}{\partial \rho}}
{\beta^2+\rho\frac{\partial\varphi}{\partial \rho}} = \frac{\beta^2 -g(\rho)}{(\beta^2+g(\rho))}
\label{L3}\ee
This equation shows the key role played by the GEM magnetic field, that  is now:
\be \frac{B_z}{c}=\frac{\rho\frac{\partial\varphi}{\partial \rho}-\beta^2}{\beta \, \rho} =\frac{g(\rho)-\beta^2}{\beta \rho}.
\label{L4}\ee
Eq.(\ref{L3}) is an elegant rate equation for the velocity outside the galaxy. However, in any phenomenology,
care must be taken to ensure that the GEM magnetic field employed ({\it vedi} Eq.(\ref{L4}) $B_z<0$ is
indeed negative. A counter example, has already been provided in Sec.(\ref{lenz}).

\section{\bf Rotation velocity and angular momentum for the Milky Way \label{phen}}
\begin{figure}
\scalebox {0.7}{\includegraphics{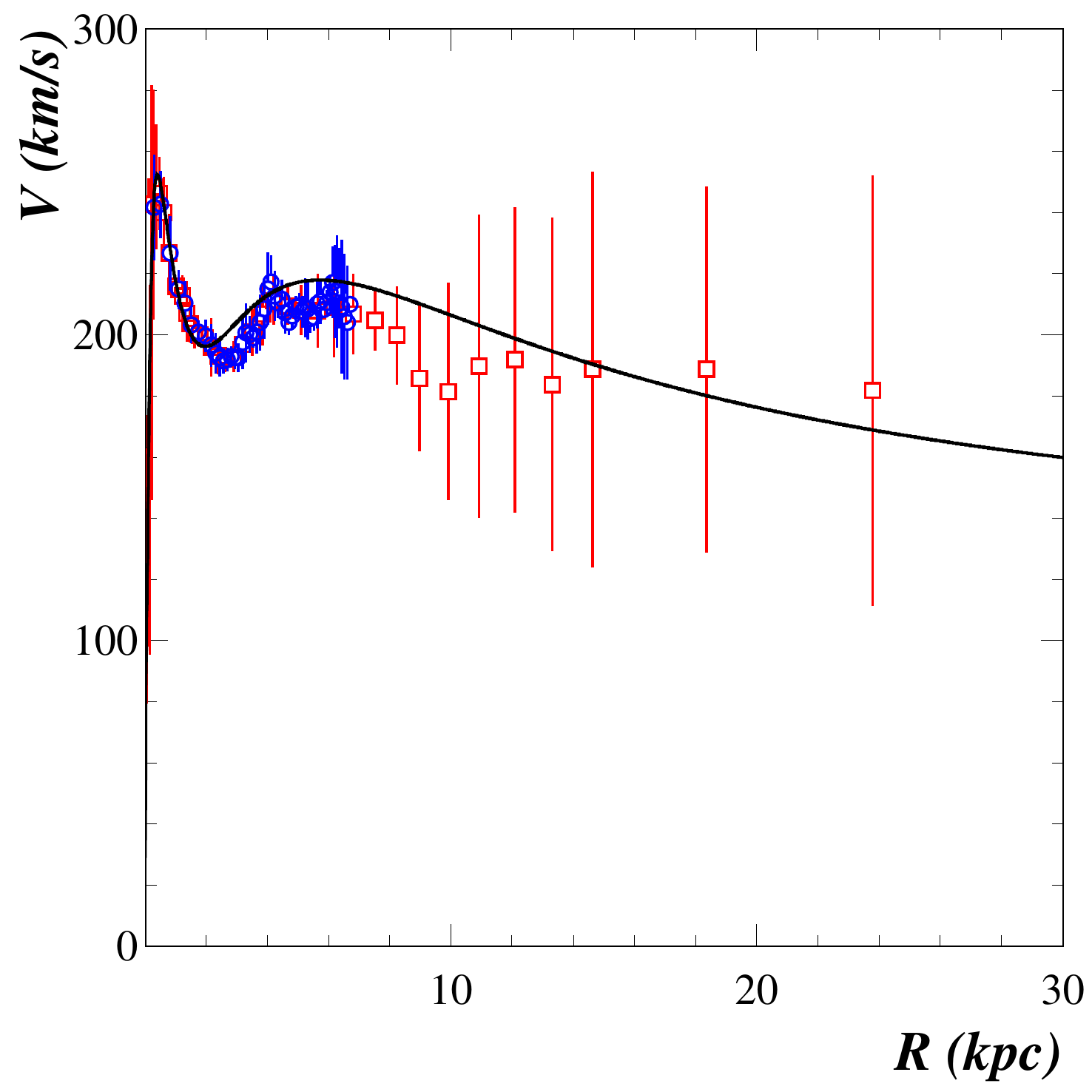}}
\caption{We show rotation velocity for our Milky Way using Eq.(\ref{p8})}
\label{Fig(4)}
\end{figure}

Our own galaxy the Milky Way is presumably the one we ought to know the best and yet it is most 
arduous to discuss it realistically given its rings and spiral arms that belie our assumption 
of axial symmetry as its structure in no way can be considered independent of the angle $\varphi$ \cite{BH}. 
In the Weyl formalism under consideration in this paper, rings and spiral arms can occur due
to instabilities generated by the motion of the interstellar medium (ISM). See, for example,\cite{Frenkler:2022}.
Following a hollowed theoretical custom, presently we shall ignore these as of no consequence 
and proceed with confidence that the Einstein theory with an extended Weyl metric and a pressure-less 
source is applicable to it and we shall be satisfied if our description is even approximately successful 
for the angular momentum and rotational velocity of this massive bar like object in terms of its known 
diameter (about 25 Kpc); thickness (about 2 Kpc) and its baryonic mass; that is, use only the {\it visible} 
part of the galaxy in trying to understand it. After all, we do feel less guilty in our maneuvers in that we 
are not assuming that our galaxy consists of a vast (over an order of magnitude more massive) amount 
of {\it unseen} dark matter (of {\it unknown} origin) spread out (over a radius of 380 Kpc) rotating with perfect 
{\it spherical symmetry} obeying {\it Newtonian} mechanics\cite{Sofue:2017}.\\
\\
To begin our phenomenology, we need an input mass density $\rho_m(\rho,z)$ that describes the bulge, the disk and a 
co-rotating gas surrounding it, a Newtonian potential and the corresponding Newtonian (squared, normalized) 
velocity $g(\rho,z=0)$ generated from it and an estimate of its baryonic mass ($M$). Unfortunately, there is less than
unanimity as to what this mass is: Allen's astronomical data lists $M_{galaxy}=1.4\times10^{11} M_\odot$
\cite{Allen}; Trimble quotes $M_{galaxy}=1\times 10^{11}M_\odot$\cite{Trimble:1984};
Nagai-Miyamoto estimate it to be about $2.567\times 10^{11}M_\odot$\cite{MN:1975}; Lipovka 
estimate is $2.3\times 10^{11}M_\odot$\cite{Lipovka:2018}; Sofue obtains for the bulge and the disk 
mass $M_{b+d}= 7.9\times 10^{10}M_\odot$, however this analysis also has a
DM halo mass of $2.23\times 10^{11}M_\odot$ (within a DM halo radius $h \sim 22$ Kpc)\cite{Sofue:2017}. It 
is important to note that the Sofue estimates include the DM component to the regular baryonic bulge and disk
 components in fitting the galaxy rotation curve at {\it small} distances. The total fraction of baryons from WMAP cosmic 
 value is $17\%$\cite{Dehnen:1998}, it is estimated to be $12\%$ as the mean for a group of galaxies, 
 whereas for our own galaxy it is only 
$5.9\%$ of DM considered spread out up to 380 Kpc (chosen arbitrarily as the half-distance between our 
and M31 galaxy nearby\cite{Sofue:2017}.\\
\\
In view of the above uncertainties, we decided to answer only the following question: Assume a baryonic mass density
$\rho_m(\rho,z)$ spread out only over the visible domain of our galaxy (roughly 25 Kpc in diameter and 2 Kpc in
thickness) whose Newtonian potential provides a reasonable description of the rotation velocity including the two 
visible bumps in the velocity along with the expected Keplerian fall-off at larger distances. We compute using the GR 
formalism described in the text: the total mass $M$ (baryonic+radiation); the total angular momentum $J$ and the 
rotation velocity. As we have stressed, the continuity constraints in GR imply that the {\it magnetic} contribution that 
keeps the velocity up at larger distances cannot be ignored since it is related to the Newtonian term. 
Thus follows the simple illustrative example.\\ 
\\     
We chose a convenient analytic (\& factorizable) mass 
distribution due to Lipovka\cite{Lipovka:2018} so as to facilitate our 
computations of the total mass, angular momentum and the Newtonian 
velocity vs. distance. The factorized mass
density reads    
\be
\label{p1}
\sigma(\rho,z)
=\frac{10^{10}M_\odot}{\lt\gamma t^2+1\rt^{3/2}}
\sum_{k=1}^n\frac{\alpha_k}{\lt\beta_k x^2+1\rt^{3/2}}
=\frac{10^{10}M_\odot}{\lt\gamma z^2/b^2+1\rt^{3/2}}
\sum_{k=1}^n\frac{\alpha_k}{\lt\beta_k \rho^2/a^2+1\rt^{3/2}}
\\
=\frac{10^{10}M_\odot a^3 b^3}{\lt\gamma z^2+b^2\rt^{3/2}}
\sum_{k=1}^n\frac{\alpha_k}{\lt\beta_k \rho^2+a^2\rt^{3/2}}\,,
\ee
where the $n$-vector parameter $\vec\beta=(\lt\beta_1,\ldots,\beta_n\rt)$ is a-dimensional, 
while the $n$-vector $\vec\alpha=(\lt\alpha_1,\ldots,\alpha_n\rt)$ has dimension [length]$^{-3}$ in units of Kpc$^{-3}$.\\ 
The mass is given by
\be
\label{p2}
M
=\int_{-\infty}^\infty dz \int_{-\infty}^\infty d\rho \,\rho \int_{0}^{2\pi} d\varphi\, 
\no\\ 
= 2\pi \int_{-\infty}^\infty dz \int_{-\infty}^\infty (d\rho\, \rho)\ \sigma(\rho,z)\,.
\ee
It can also be written as
\be
\label{p3}
M
=4\pi10^{10}M_{\odot} \sum_{k=1}^n \frac{b\alpha_k/\sqrt{\gamma +1}}{\beta_k/a^2}\lt [1- \frac{1}{ \sqrt{\beta_k+1}}]\rt
\\
=4\pi10^{10}M_{\odot}\sum_{k=1}^n \frac{\alpha^*_k}{ \beta_k^*}\lt [1- \frac{1}{ \sqrt{\beta_k^*a^2+1}}]\rt\,,
\ee
where the asterisk parameters are
\be
\label{p4}
\alpha_k^*=\frac{b\alpha_k}{\sqrt{\gamma+1}}
\,,\hh\hh
\beta_k^*=\frac{\beta_k}{a^2}\,,\hh\hh k=1,2,\ldots,n\,.
\ee
Using the following values
\be
\label{p5}
a=12.5\,\mbox{kpc}
\,,\hh
\vec\alpha^*=(\alpha_1^*,\alpha_2^*,\alpha_3^*)
=\lt 0.2317, 6.358, 7.005\rt\,\mbox{kpc}^{-2}
\,,\hh
\vec\beta^*=(\beta_1^*,\beta_2^*,\beta_3^*)
=\lt 0.112, 28.8, 1440\rt\,\mbox{kpc}^{-2}
\ee
The above equations give
\be
\label{p6}
M\simeq \lt 2.37\times 10^{11}\rt\,M_{\odot}
\ee
By considering the first set of Lipovka's parameters, namely
\be
a=12.5\,\mbox{kpc}\,,\hh\hh
\vec\alpha=(\alpha_1,\alpha_2)=(0.24,3.7)\,\mbox{kpc}^{-3}\,,\hh\hh
\vec\beta=(\beta_1,\beta_2)=(12,10000)\,,\hh\hh
\gamma=30\,,
\nn\ee 
we have to use the mass formula in terms of the non-asterisk parameters, namely
\be
M=4\pi10^{10}M_{\odot}\frac{a^2b}{\sqrt{\gamma+1}}
\sum _{k=1}^2\frac{\alpha_k}{\beta_k}\lt[1-\frac{1}{\sqrt{\beta_k+1}}]\rt\,,
\nn\ee
and one needs to know the Galaxy half-thickness $b$, together with its characteristic radius $a$
 Please note that the asterisk parameters embody such information, in particular, the half-thickness 
 $b$ enters the definition of the $n$-vector $\vec\alpha^*$ in terms of $\vec\alpha$. So that, as a 
 function of the ratio $s=b/a$ we have
\be
M
= 4\pi10^{10}M_{\odot}\frac{s a^3}{\sqrt{\gamma+1}}
\sum _{k=1}^2\frac{\alpha_k}{\beta_k}\lt[1-\frac{1}{\sqrt{\beta_k+1}}]\rt
\simeq s (\lt 1.38\times 10^{11})\rt\,M_\odot
\nn\ee
The expressions for the velocity squared in terms of the non-asterisk and asterisk parameters, as a function of $\rho$ are respectively 
\be
\label{p7}
V_\perp^2(\rho)      
=2\pi 10^{10}M_\odot G\frac{s \,a^3}{\rho\sqrt{\gamma+1}}
\sum_{k=1}^n\frac{\alpha_k}{\beta_k}\lt[1-\frac{(3/2)\beta_k \rho^2/a^2+1}{\lt \beta_k \rho^2/a^2+1\rt^{3/2}}]\rt\,,
\no\\
=\frac{2\pi 10^{10}M_\odot G}{\rho}
\sum_{k=1}^n\frac{\alpha_k^*}{\beta_k^*}\lt[1-\frac{(3/2)\beta_k^* \rho^2+1}{\lt \beta_k^* \rho^2+1\rt^{3/2}}]\rt\,.
\ee
The geo-magnetic velocity  has been chosen to asymptote to a constant as discussed in the text. In units of Kpc,
it reads
\be
\label{p8}
V_{\rm mag}(\rho)=(\frac{160}{3.09\times 10^{16}})[\frac{\rho}{\rho+70}]\,,\ee
where $\rho$ is in units of Kpc.\\
The modified velocity is
\be
\label{p9}
V_{\rm mod}(\rho)=\frac{V_{\rm mag}(\rho)+\sqrt{4V_\perp^2(\rho)+V^2_{\rm mag}(\rho)}}{2}\,.
\ee
Fig(4) shows the modified velocity using Eqs.(\ref{p7},\ref{p8} \&\ \ref{p9})
with $n=3$.
\\
\\
The non-modified and the modified total angular momentum are given by
\be
\label{p10}
J= 2\pi\int_0^a \rho d\rho\int_{-b}^b dz\, \sigma(\rho,z)V_\perp(\rho)\rho;\ \&\ \  
J_{\rm mod}= 2\pi\int_0^a \rho d\rho\int_{-b}^b dz\, \sigma(\rho,z)V_{\rm mod}(\rho)\rho\,,
\ee
respectively. The mass density given in Eq.(\ref{p1}) depends on the half-thickness $b$, or in the 
ratio $s=b/a$, however in computing the total angular momentum such a dependency is canceled out 
by the integration. The non-modified and modified total angular momenta are
\begin{eqnarray}
\label{p11}
J\simeq 1.155\times 10^{67}\ \ {\rm Joules-sec} \ \  J_{\rm mod}\simeq 1.193\times 10^{67}\ {\rm Joules-sec}.
\end{eqnarray}
This estimate can be compared to Trimble's estimate\cite{Trimble:1984} of the angular 
momentum $6\times 10^{66}$ Joules-sec, obtained using {\it only the disk} part of the Milky Way.

\section{\bf Conclusions \& future prospects \label{conc}}
\noindent
Here we first summarize results obtained, then describe research in progress 
and close with prospects for the future.\\
\begin{itemize}
\item 1. Our work began with the most general framework in GR to discuss rotationally supported 
galaxies. Fortunately, there is the Weyl class of axisymmetric metrics for whom the solutions to the Einstein-Weyl 
equations in the vacuum are known in terms of a few differential equations. Even more fortunately, for what we call 
the extended Weyl class that includes rotations explicitly, exact differential equations are also known; 
\item 2. Unlike the Kerr metric, Weyl metric can be easily (and has been) continued within the galaxy and physically 
meaningful results obtained;
\item 3. Armed with exact solutions, it became possible to show how Gau\ss\ and Amp\'ere laws emerged and under
 what conditions  Ludwig's extended GEM theory and his  non-linear rate equations for the rotation velocity field could be deduced;
\item 4. Using the century old iterative procedure in GR and further elaborated by Weinberg, we could discuss
the value of the mass M (baryonic mass +that of the gravitational field) \&\ that of the intrinsic angular 
momentum $J$ of a rotationally-supported galaxy. The extended Weyl metric analysis allowed us to conclude 
rigorously that Weyl's (vectorial) length parameter $a$ must have a finite limit to obtain a finite $J$. As the same 
parameter also controls the asymptotic limit of the rotation velocity, we can conclude that GR is indeed capable of
 obtaining a flat plateau in the rotation velocity.
\item 5. We have attempted an alternative strategy to that of Ludwig as far as the phenomenology of the rotation curves
is concerned. Ludwig eliminated the magnetic contribution to obtain his non-linear rate equation for the velocity field
 in terms of the input from the Newtonian potential and the mass distribution within the galaxy. Instead, we kept the 
 Newtonian input \&\ the magnetic input together -thus our velocity equations remained linear. This allowed us to provide 
 a clear physical picture: at small distances, the velocity is basically described by the Newtonian term and as it 
 begins to fall off it is supported near the edge by essentially a constant vector potential. It also brought to focus the 
 crucial role of Lenz's law and the left hand rule for the GEM magnetic field.
\item 6. As by products of our analysis, we have deduced a few other practical results: (i) Imposition of Lenz's law 
implies the rigorous inequality: $\beta^2\geq g$, the Newtonian value. A result supported by 2700 data points from 
153 rotating galaxies;  (ii) a better estimate ($\geq 500$ Km/sec.) for our Sun's escape velocity from our galaxy; (iii) an 
easy to remember pneumonic for the asymptotic velocity $\beta^2\approx(R_s/(2R_{edge}))$; (iv) how Tully-Fisher 
law emerges from a rotating {\it pill-box} galaxy; (v) Simple dimensional analysis implies $J\propto M^{7/4}$ if Tully-Fischer holds. 
\end{itemize}
Our present focus is four fold:
\\
A: A satisfactory GR description of the deflection of light from large galaxies \&\ from galaxy clusters;\\
\\
B: To obtain a better understanding of the TF-law ($M\propto \beta^4$) and the Virginia Trimble law ($J\propto M^{1.9}$),
the latter covering data that run over 50 orders of magnitude\cite{Trimble:1984}.\\
\\
C: A comprehensive phenomenolgy of the rotation curves with realistic densities and more refined Newtonian inputs.\\
\\
D: Testing our conjecture that spiral arms in rotating galaxies such as ours are generated dynamically through non-linear 
effects inherent in GR.\\
\\
On the broader horizon, it is reasonable to hope further yet more brilliant advances in astrophysical observations (for
example, via renewed investigations involving Hanbury-Brown-Twiss techniques) so as to reduce the error bars in 
rotation curves. Only then, it would be possible to truly distinguish between different theoretical models. 

\section{\bf Acknowledgements}
YS would like to thank Professor Gerson Ludwig for numerous correspondence and discussions about his pioneering 
 work on  rotationally supported galaxies. He would also like to thank members of the dipartimento di fisica e geologia 
 di Universit\'a di Perugia for their hospitality \& to Dr. Patrizia Cenci, Direttore di INFN, Sezione di Perugia for her continued
 encouragement and warm support.   

\section{Appendix A: Exact non-linear expression for the velocity field \label{AppA}}
The exact expressions for $a_{,\rho}$ \&\ $a_{,z}$ read
\begin{eqnarray}
\label{A1}
\frac{a,\rho}{\rho}= -(\frac{w e^{-4U}}{\rho}) +(\frac{1}{w}+we^{-4U})U_{,\rho};\nonumber\\
 \frac{a,z}{\rho}=  +(\frac{1}{w}+w e^{-4U}) U_{,z};\nonumber\\
 {\rm where}:\ w =\frac{\beta}{1-\beta(\frac{a}{\rho})}
\end{eqnarray}
Following exactly the steps described in Eqs.(\ref{a3}-\ref{a5}) {\it et sec.} in Sec.(\ref{matter}), we find
two expressions for [$U_{,\rho ,\rho}+U_{,z,z}$], which we equate and find
\begin{eqnarray}
\label{A2}
 -\frac{U_{,\rho}}{\rho}-(\frac{e^{4U}}{2})\big{[}(\frac{1}{w}+we^{-4U})^2 U_{,z}^2 + [ -\frac{w e^{-4U}}{\rho} +
  (\frac{1}{w}+we^{-4U}) U_{,\rho}]^2 \big{]} \nonumber\\
=  (\frac{we^{-4U}}{1+w^2e^{-4U}})(\frac{w}{\rho})_{,\rho}- (ln[\frac{e^{4U}}{w}+w])_{,\rho}) U_{,\rho} 
-(ln[\frac{e^{4U}}{w}+w])_{,z}) U_{,z};\nonumber\\
\end{eqnarray}
Once again, on the equatorial plane $z=0$, using the up-down symmetry, we can drop all terms 
such as $w_{,z}$ and $U_{,z}$ and thus remaining with
 \begin{eqnarray}
\label{A3}
 -\frac{U_{,\rho}}{\rho}-(\frac{e^{4U}}{2})[ -\frac{w e^{-4U}}{\rho} + (\frac{1}{w}+we^{-4U}) U_{,\rho}]^2 \nonumber\\
=  (\frac{we^{-4U}}{1+w^2e^{-4U}})(\frac{w}{\rho})_{,\rho} -(ln[\frac{e^{4U}}{w}+w])_{,\rho}) U_{,\rho} 
;\nonumber\\
(\frac{w^2 e^{-4U}}{\rho})U_{,\rho} +\frac{e^{-2U}\beta^2}{w^2\gamma^2}
=(\frac{we^{-4U}}{1+w^2e^{-4U}})(\frac{w_{,\rho}}{\rho})
-(ln[\frac{e^{4U}}{w}+w])_{,\rho}) U_{,\rho}+(\frac{e^{4U}}{2})(\frac{1}{w}+we^{-4U})^2 U_{,\rho}^2;\nonumber\\
\end{eqnarray}
\\
\section{\bf Appendix B: Iterative  computational procedure in GR \label{iter}}
\noindent
During the past century, a detailed program [often dubbed, Post- Newtonian, Post-post Newtonian 
etc.] was developed to systematically compute the metric, the Ricci tensor and the like in a perturbation 
expansion in powers of the Newton's constant $G$. The procedure is somewhat involved but technically 
straightforward albeit cumbersome. And it does require the introduction of a non-tensorial object first introduced 
by Einstein and called by him the pseudo-energy momentum tensor for the gravitational field. It was formalized by 
Landau \& Lifshitz\cite{LL:1965} and is amply discussed in the excellent textbooks such as that by 
Weinberg\cite{Weinberg:1972} and by Stephani\cite{Stephani:1990}. In order not to duplicate some long
 expressions, we shall refer the reader to these references abbreviated as (L\&L),W or S.\\
A few words are in order as to the reason for this Appendix. While well known to physicists of the last generation, 
our own experience has been that the detailed procedures are largely forgotten by a vast majority of practicing physicists. 
Thus, to bring out the differences with the traditional post-Newtonian theory and to stress the
 importance of what is involved in the very definition of the {\it far field}, we here review the iterative formalism in some detail. Another point to stress here is that the exact Weyl solutions for the vacuum that are discussed here appear to be analytically continuable within the system (say a galaxy) and exchanges of energy-momentum emerge at order $G^2$. Thus, a diligent reader can compare for herself the exact results with pieces constructed from higher order iterative solutions.  \\
\\
Consider the Einstein equation with its prescribed source, a matter energy-momentum tensor that is limited in its spatial and temporal extent.
\begin{eqnarray}
\label{I1}
G_{\mu\nu}\equiv R_{\mu\nu}- \frac{1}{2} g_{\mu\nu} R= (\frac{8\pi G}{c^4}) T_{\mu\nu};\ R_{\mu\nu}=(\frac{8\pi G}{c^4}) [T_{\mu\nu}- \frac{1}{2} g_{\mu\nu} T];\    ; (i)\nonumber\\
 T^\nu_{\ \mu;\nu} = \frac{1}{\sqrt{-g}}\frac{\partial (T^\nu_{\ \mu} \sqrt{-g})}{\partial x^\nu}-\frac{1}{2}(\frac{\partial g_{\nu\lambda}}{\partial x^\mu}) T^{\nu\lambda} =0; (ii)\nonumber\\
R_{\mu\nu} = R_{\mu\nu}^{(1)} + R_{\mu\nu}^{(2)}; (iii)\nonumber\\
R_{\mu\nu}^{(1)}= \frac{\partial \Gamma^\lambda_{{\mu\nu}}}{\partial x^\lambda} -\frac{\partial \Gamma^\lambda_{{\mu\lambda}}}{\partial x^\nu}= \frac{1}{2}g^{\lambda\sigma}\big{[} \frac{\partial^2g_{\mu\lambda}}{\partial x^\nu \partial x^\sigma}+ \frac{\partial^2g_{\nu\lambda}}{\partial x^\mu \partial x^\sigma}- \frac{\partial^2g_{\lambda\sigma}}{\partial x^\mu \partial x^\nu}- \frac{\partial^2g_{\mu\nu}}{\partial x^\lambda \partial x^\sigma} \big{]}
; (iv)\nonumber\\
R_{\mu\nu}^{(2)}= \big{[}\Gamma^\lambda_{\mu\nu}\Gamma^\sigma{\lambda\sigma}-\Gamma^\sigma_{\mu\lambda}\Gamma^\lambda_{\nu\sigma}\big{]}; (v)\nonumber\\
\end{eqnarray}
Note that the vanishing of the covariant divergence of the (material) energy-momentum tensor $T^\nu_{\ \mu}$ as given in Eq.(\ref{I1}(ii)) does not lead to a local energy-momentum or, -angular momentum- conservation law. This reflects the physical fact that in a gravitational field the 4-momentum of the matter field alone is not conserved, but rather the 4-momentum of matter plus that of the gravitational field; the latter is not included in $T^\nu_{\ \mu}$. Thus, one defines a pseudo energy-momentum tensor $t^{\mu\nu}$ for the gravitational field\cite{QCD}, so that the following condition holds:
\begin{eqnarray}
\label{I2}
\partial_\nu (T^{\mu\nu}+t^{\mu\nu})=0;
\end{eqnarray}
We know that $t^{\mu\nu}$ is not a tensor, ordinary derivative in Eq.(\ref{I2}) confirms this fact. However, we can devise recipe so that asymptotically the fields are Lorentz covariant. Below are the steps of the perturbative recipe:\\
{\bf Step I:}\\ 
\noindent
We know that there exists a space-time point at which all the $\Gamma$'s can be made to vanish (the first derivatives of the metric but not the metric itself can be made to vanish). But this implies\\ 
(a) through Eq.(\ref{I1}(ii)) that the last term disappears; that the determinant of the metric can be taken out of the partial derivative in the first term, rendering the covariant derivative to an ordinary derivative, i.e., $\partial_\nu T^{\mu\nu}=0$ at this point;\\
(b) simultaneously, we learn from Eq.(\ref{I1}(v)) that $R_{\mu\nu}^{(2)}$ vanishes at this point. Thus the entire Einstein Eq(\ref{I1}(i)) is reduced (at this space-time point) to 
\begin{eqnarray}
\label{I3}
G_{\mu\nu}\to G_{\mu\nu}^{(1)}= R_{\mu\nu}^{(1)}- \frac{1}{2} g_{\mu\nu} R^{(1)}= (\frac{8\pi G}{c^4}) T_{\mu\nu};\nonumber\\
\end{eqnarray}
Consider the special case (certainly valid for weak-gravity) that the metric can be expanded around its flat Minkowski limit $\eta_{\mu\nu}$ and for computational simplicity choose {\it harmonic coordinates}. (Indices being raised and lowered by $\eta_{\mu\nu}$).
\begin{eqnarray}
\label{I4}
{\rm harmonic\ coordinates}: g^{\mu\nu}\Gamma^\lambda_{\mu\nu}=0;\ {\rm picking\ the\ gauge:\ coordinate\ conditions};(i) \nonumber\\
g_{\mu\nu}= \eta_{\mu\nu}+h_{\mu\nu};\  \bar{h}_{\mu\nu}= h_{\mu\nu}-(\frac{1}{2})h\eta_{\mu\nu};\ h=h_\mu^\mu=-\bar{h};\nonumber\\
g_{\mu\nu}= \eta_{\mu\nu}+\bar{h}_{\mu\nu}-(\frac{1}{2})\eta_{\mu\nu}\bar{h};\nonumber\\
{\rm So}:\ R_{\mu\nu}^{(1)}= \frac{1}{2}(\bar{h}^\lambda_{\ \mu,\nu,\lambda}
+\bar{h}^\lambda_{\ \nu ,\mu,\lambda} -(\partial_\lambda\partial^\lambda)\bar{h}_{\mu\nu}+
\frac{1}{2}\eta_{\mu\nu}(\partial_\lambda\partial^\lambda)\bar{h});\nonumber\\
{\rm And}: G_{\mu\nu}^{(1)}= R_{\mu\nu}^{(1)}-\frac{1}{2}\eta_{\mu\nu}R^{(1)}=
 -\frac{1}{2}(\partial_\lambda\partial^\lambda)\bar{h}_{\mu\nu} + \mathcal{X}_{\mu\nu};\nonumber\\
{\rm where}\ \mathcal{X}_{\mu\nu}= \frac{1}{2}\big{(} \bar{h}^\lambda_{\ \mu,\nu,\lambda}+
\bar{h}^\lambda_{\ \nu,\mu,\lambda} -\eta_{\mu\nu}\bar{h}^{\lambda\sigma}_{\ \ ,\lambda ,\sigma} \big{)}
\end{eqnarray}
We can eliminate $\mathcal{X}_{\mu\nu}$ through the following 4-coordinate condition choice allowed by Eq.(\ref{I4}(i)) (Details can be checked via Eqs.(Stephani13.8-13.14)):
\begin{eqnarray}
\label{I5}
{\rm harmonic\ coordinates\ defined\ by}:\ {\rm curvilinear\ D'Alembertian}(x^\mu)= \frac{1}{\sqrt{-g}}(\sqrt{-g}\ g^{\nu\lambda} x^\mu,\lambda)_{,\nu}=0;(i)\nonumber\\
{\rm Implies}\ (\sqrt{-g}\ g^{\mu\nu})_{,\nu}=0;(ii)\nonumber\\
{\rm Thus,\ if}\  \sqrt{-g}\ g^{\mu\nu} = \eta^{\mu\nu}- \tilde{\bar{h}}^{\mu\nu},\ {\rm we\ have}\ \tilde{\bar{h}}^{\mu\nu}_{\ \  ,\nu}=0;(iii) \nonumber\\
\end{eqnarray}
To accomplish it, we need to make a coordinate change as follows:
\begin{eqnarray}
\label{I6}
{\rm Let}\  \bar{x}^\mu= x^\mu+b^\mu;\nonumber\\
{\rm implying\ a\ change\ in\ the\ metric}\ \bar{g}^{\mu\nu}=g^{\lambda\sigma}(\delta^\mu_\lambda+b^\mu_{\ ,\lambda})(\delta^\nu_\sigma+b^\nu_{\ ,\sigma})\approx\ g^{\mu\nu}+g^{\lambda\nu}b^\mu_{\ ,\lambda}+g^{\lambda\mu}b^\nu_{\ ,\lambda};\nonumber\\
\&\  {\rm a\ change\ in\ the\ metric\ determinant}\ \bar{g}=|\bar{g}^{\mu\nu}|^{-1}\approx\ g(1+2b^\lambda_{\ ,\lambda})^{-1}
;\nonumber\\
\tilde{\bar{h}}^{\mu\nu}=\bar{h}^{\mu\nu}- b^{\mu\ ,\nu}-b^{\nu\ ,\mu}+\eta^{\mu\nu}b^\lambda_{\ ,\lambda};\nonumber\\
{\rm Imposing}\ \bar{h}^{\mu\nu}_{\ \ ,\nu}= (\partial_\sigma\partial^\sigma)b^\mu;\nonumber\\
{\rm so\ that}\ G_{\mu\nu}^{(1)}\to (-\frac{1}{2})(\partial_\sigma\partial^\sigma)[\bar{h}_{\mu\nu}-b_{\mu\ ,\nu}-b_{\nu\ ,\mu}+\eta_{\mu\nu}b^\lambda_{\ ,\lambda}]=(-\frac{1}{2})(\partial_\sigma\partial^\sigma)\tilde{\bar{h}}_{\mu\nu};\nonumber\\
{\rm And}\ \ \ \tilde{\bar {h}}^{\mu\nu}_{\ \ \ ,\nu}=0;\nonumber\\
\end{eqnarray}
by virtue of Eq.(\ref{I5}(iii)). This completes the proof and we have (dropping the tilde and the bar on h), in harmonic co-ordinates: 
\begin{eqnarray}
\label{I7}
G_{\mu\nu}^{(1)}= (-\frac{1}{2})(\partial_\sigma\partial^\sigma)h_{\mu\nu}=(\frac{8\pi G}{c^4})T_{\mu\nu};\ (i)
\nonumber\\
(\partial_\sigma\partial^\sigma)h_{\mu\nu}=(-\frac{16\pi G}{c^4})T_{\mu\nu};(ii)\nonumber
\end{eqnarray}
\\
\\
{\bf Step II:}
\noindent
By definition
\begin{eqnarray}
\label{II-1}
G_{\mu\nu}^{(1)} = (\frac{8\pi G}{c^4})[T_{\mu\nu}+t_{\mu\nu}];\nonumber\\
{\rm (i)\ where\ the\ gravitational\ energy-momentum\ tensor\ is\ given\ by}\nonumber\\ 
t_{\mu\nu}= -(\frac{c^4}{8\pi G})[G_{\mu\nu}-G_{\mu\nu}^{(1)}];\nonumber\\
{\rm (ii)\ the\ total\ matter+\ gravitational\ energy-momentum\ tensor}\nonumber\\ 
\tau^{\lambda\sigma}\equiv \eta^{\lambda\mu}\eta^{\sigma\nu}[T_{\mu\nu}+t_{\mu\nu}];\nonumber\\
{\rm that\ is\ locally\ conserved}\ \tau^{\lambda\sigma}_{\ \ ,\sigma}=0;\nonumber\\
\end{eqnarray}
Let us note here the convention that indices for quantities such as $h_{\mu\nu};G_{\mu\nu}^{(1)}$ and $\frac{\partial}{\partial x^\lambda}$ are raised \& lowered by the $\eta$'s, whereas on true tensors such as $R_{\mu\nu}$ are raised and lowered with g's as usual.\\
Weinberg Chapter 7(Sec. 6) describes in detail the definition of total momentum, total energy \& the angular momentum, as well as the computational strategy for a perturbative expansion of $t_{\mu\nu}$. We list some of them below for reference: 
\begin{eqnarray}
\label{II-2}
{\rm Total\ 4-momentum}\ P^\mu= \int_{V} \tau^{o\mu} (d^3x);\&\ \tau^{i\nu}\ {\rm is\ the\ flux};\nonumber\\
{\rm Total\ Angular-momentum\ density\ and\ flux}\ M^{\mu\nu\lambda}= \tau^{\mu\lambda}x^\nu-\tau^{\mu\nu}x^\lambda;
\nonumber\\ 
\partial_\mu M^{\mu\nu\lambda}=0;\ {\rm since}\ \tau^{\nu\lambda}= \tau^{\lambda \nu}\ \&\ \tau^{\mu\nu}_{\ \ \ ,\nu}=0;\nonumber\\
{\rm Total\ Angular-Momentum}\ J^{\nu\lambda}= -J^{\lambda\nu}= \int_{V}(d^3x)M^{o\nu\lambda}\ ({\rm a\ constant\ if\ no\ surface\ terms});\nonumber\\
\end{eqnarray}
In Weinberg(Eqs.(7.6.14-15)), a power series for $t_{\mu\nu}$ in h is developed up to terms of order $h^2$:
\begin{eqnarray}
\label{II-3}
t_{\mu\nu}=(\frac{c^4}{8\pi G}) [-(\frac{1}{2})h_{\mu\nu} R^{(1)\,\lambda}_{\ \ \lambda}
+\frac{1}{2}\eta_{\mu\nu}h^{\lambda\sigma}R^{(1)}_{\lambda\sigma}
+R^{(2)}_{\mu\nu}-\frac{1}{2}\eta_{\mu\nu}\eta^{\lambda\sigma} R^{(2)}_{\lambda\sigma}]+  
\bigcirc(h^3);\nonumber\\
\end{eqnarray}
where $R^{(2)}$ is given by the terms of order $h^2$ in Eq.(\ref{I1}(v)) and are written out in detail in Weinberg(Eq.(7.6.15)).\\
\\
Far away from the finite material system that produces the gravitational field, $T^{\mu\nu}$ vanishes and since, $t_{\mu\nu}$ is of order $h^2$, the source terms on the rhs of Eq.(\ref{II-1}) are confined to a finite region. Thus, we expect them to behave as electrostatic potentials or as in Newtonian gravitational theory. Typically, we expect for large distances from the source that
\begin{eqnarray}
\label{II-4}
h_{\mu\nu}\to \bigcirc(\frac{1}{r});\ \frac{\partial h_{\mu\nu}}{\partial x^\lambda}\to \bigcirc(\frac{1}{r^2});\ \frac{\partial h_{\mu\nu}}{\partial x^\lambda\partial x^\sigma}\to \bigcirc(\frac{1}{r^3});\nonumber\\
{\rm Hence}\  t_{\mu\nu}\to \bigcirc({\frac{1}{r^4}});\ {\rm by\ virtue\ of} Eq.(\ref{II-3});\nonumber\\
\end{eqnarray}
so that the integrals for the total momentum \& energy as given in Eqs(\ref{II-2}) should converge. In fact, there are very simple expressions for the total energy and the angular momentum of a finite system(to linear order in the metric perturbations) :
\begin{eqnarray}
\label{II-5}
E_{Total}=P^oc = -(\frac{c^4}{16\pi G})\int \big{[}\frac{\partial h_{jj}}{\partial x^i} - \frac{\partial h_{ij}}{\partial x^j} \big{]})(n_i r^2d\Omega);\nonumber\\
J^{jk}= -(\frac{c^3}{16\pi G})\int K_{ijk} (n_ir^2d\Omega);\ n_i\ {\rm is\ the\ outward\ normal}\ ;\nonumber\\
K_{ijk}= (-x_j h_{ok,i}+x_k h_{oj,i})+(x_j h_{ki,o}-x_k h_{ji,o})+(h_{ok}\delta_{ij}-h_{oj}\delta_{ik});\ J_1=J^{23}, etc.
\end{eqnarray}
For the above computations, only the asymptotic behavior of the metric is required (at large distances from the source). It should also be noted, that while the total energy can be proven to be positive (provided there is a mass in the system), total angular momentum is strictly zero unless (asymptotically) either (i) the purely spatial metric is time-dependent or (ii) there is a non-trivial (i.e.,non-removable by a coordinate transformation) $h_{oi}$. As important examples, one finds by explicit calculation that both for the Schwarzschild metric and the Kerr metric that the total energy $E_{total}=Mc^2$. On the other hand, the total angular-momentum for the Schwarzschild case is zero, whereas for the Kerr metric, $J=(Mc)a$, where a is a length parameter associated with rotations.  
\\  
Of course, as Weinberg explicitly cautions, Eq.(\ref{II-4}) need not be true always. He gives the standard example of a system that has been continuously radiating energy (as gravitational waves) and so the total energy is indeed infinite: It shows up theoretically in that various derivatives become all of the same order violating Eq.(\ref{II-4}).\\
While the above is evidently acceptable on physics grounds, there are other more subtle effects that can invalidate or certainly modify the {\it reasonable} sounding estimates provided by Eq.(\ref{II-4}) augmented by our deep seated  Newtonian bias. One of them concerns rotations\cite{Pietronero:1973}. Simply because rotation about a fixed axis differentiates between clockwise \& anti-clockwise motion. Suppose the rotation is about the z-axis confined to the ($x-y$) plane and if the system is axially symmetric, $J_z$ would be conserved. It would appear that $PT$ would be conserved but not $P$ or $T$ separately, because by assumption our system is rotating with respect to an external inertial observer. For rotations that are measurable in the {\it far field}, traditional power counting methods need to be critically examined.\\ 
For the problem at hand i.e., the dynamics of rotation-supported galaxies, it is obviously not only convenient but appears mandatory that the kernel of {\it perturbative solution} include not only the Newtonian potential $U$ but also Weyl's rotation field $a$ explicitly. Technically this means that the ``asymptotic'' metric not be Galilean but augmented by the Weyl field in such a manner that a finite total angular momentum of the system is simply reproduced.  
\\


\begin{thebibliography}{50}
\bibitem{BH}
To say nothing of a massive black hole of mass ($4\times 10^6 M_\odot$) and a Schwarzschild radius 
($1.2\times 10^7$ Km) that exists at the center of the galaxy.
\end{thebibliography}
\end{document}